\documentclass[preprint]{elsarticle}

\usepackage{amsmath,amsfonts,amssymb}
\usepackage{mathrsfs} 
\usepackage{enumerate}
\usepackage{fullpage}
\usepackage{float}
\usepackage{xcolor}
\usepackage{graphicx}
\usepackage{caption}
\usepackage{subcaption}
 \usepackage{bmpsize}
\usepackage{framed}
\usepackage{ntheorem}
\usepackage{url}

\PassOptionsToPackage{table}{xcolor}
\PassOptionsToPackage{table}{color}

\newcommand{\reals}{\mathbb{R}}

\newcommand{\mb}[1]{\mathbf{#1}}

\newcommand{\argmin}{\arg\!\min}

\newcommand{\diag}{\textbf{diag}}

\newcommand{\myunits}[1]{\mathrm{#1}}
\newcommand{\mybunits}[1]{\myunits{[#1]}}

\newtheoremstyle{assumptionstyle}% name of the style to be used
  {}% measure of space to leave above the theorem. E.g.: 3pt
  {}% measure of space to leave below the theorem. E.g.: 3pt
  {}% name of font to use in the body of the theorem
  {}% measure of space to indent
  {}% name of head font
  {}% punctuation between head and body
  {}% space after theorem head; " " = normal interword space
  {}% Manually specify head

\theoremstyle{break}

\newtheorem*{assumption-non}{\bf{Assumptions}}

\begin{document}

\begin{frontmatter}

%\title{Estimating fugitive emissions of airborne particulate matter 
% using a Gaussian plume model} 
\title{Bayesian estimation of airborne fugitive emissions \\
 using a Gaussian plume model}

\author[sfu]{Bamdad Hosseini\corref{cor1}}
\ead{bhossein@sfu.ca}
%\ead[url]{http://www.math.sfu.ca/~stockie}

\author[sfu]{John M. Stockie}
\ead{stockie@math.sfu.ca}
\ead[url]{http://www.math.sfu.ca/~stockie}

\cortext[cor1]{Corresponding author} 
\address[sfu]{Department of Mathematics, Simon Fraser University, 8888
  University Drive, Burnaby, BC, V5A 1S6, Canada}

\begin{abstract}
  A new method is proposed for estimating the rate of fugitive emissions
  of particulate matter from multiple time-dependent sources via
  measurements of deposition and concentration. We cast this source
  inversion problem within the Bayesian framework, and use a forward
  model based on a Gaussian plume solution.  We present three alternate
  models for constructing the prior distribution on the emission rates
  as functions of time. Next, we present an industrial case study in
  which our framework is applied to estimate the rate of fugitive
  emissions of lead particulates from a smelter in Trail, British
  Columbia, Canada.  The Bayesian framework not only provides an
  approximate solution to the inverse problem, but also quantifies the
  uncertainty in the solution. Using this information we perform an
  uncertainty propagation study in order to assess the impact of the
  estimated sources on the area surrounding the industrial site.
\end{abstract}

\end{frontmatter}

%%%%%%%%%%%%%%%%%%%%%%%%%%%%%%%%%%%%%%%%%%%%%%%%%%%%%%%%%%%%%%%%%%%%%%%%
%%%%%%%%%%%%%%%%%%%%%%%%%%%%%%%%%%%%%%%%%%%%%%%%%%%%%%%%%%%%%%%%%%%%%%%%
%\linenumbers 
%\setpagewiselinenumbers
\section{Introduction} 
\label{sec:intro}

Dispersion of airborne pollutants emitted from anthropogenic sources and
their effect on the surrounding environment have been a growing source
of concern over the past several decades.  Both primary polluters and
government monitoring agencies dedicate significant resources to
tracking and controlling the release of particulate emissions from
industrial operations. Atmospheric dispersion modelling, which is the
study of mathematical models and numerical algorithms for simulating
processes involved in dispersion of pollutants from a known source, is a
vital tool for monitoring of existing industrial operations as well as
assessing the potential risk and impact of future operations.  Many
dispersion models address the situation where a pollutant source has
already been identified and the source emission rate is known
approximately, and many industry standard software packages such as
AERMOD~\cite{aermod} and CALPUFF~\cite{calpuff} are already available to
solve this problem.

In many practical settings the main problem of interest is not to
determine the impact of known sources but rather to solve the source
identification problem, which refers to determining the emission
properties and possibly also locations for a collection of unknown
sources.  Inverse source identification is particularly prominent in the
study of fugitive sources, which are particulate or gaseous emissions
that derive from leaks or other unintended releases from building
windows or vents, entrainment from debris piles, or operations of trucks
and loading equipment.  In these situations, it is usually not possible
to obtain direct measurements of the fugitive source emissions by
installing sensors; this is in contrast with emissions from stacks and
exhaust vents where such measurements are routine.  Nevertheless, it is
often still possible to take \emph{indirect measurements} of fugitive
emissions, for example by measuring concentration of a pollutant at a
remote location some distance from the source.  In this case, the
challenge is to estimate the rate of emissions based on indirect
measurements, which can be posted as an inverse problem~\cite{isakov,
  kabanikhin}.  

Atmospheric dispersion modelling is a well-developed area of research,
and for a comprehensive overview we refer the reader to the work of
Zlatev~\cite{zlatev} or Dimov et al.~\cite{dimov}.  The use of partial
differential equation (PDE) models based on the advection-diffusion
equation for modelling short-range transport of pollutants dates back to
the work of Taylor~\cite{taylor1915eddy}, Roberts~\cite{roberts-1924}
and Sutton~\cite{sutton1932theory}.  In simple scenarios involving
constant emissions, steady-state transport, point or line sources, or
flat topography, it is possible to derive approximate analytical
solutions to the governing equations~\cite{arya, okamoto-etal-2001,
  park-baik-2008, seinfeld}.  These analytical solutions, referred to
collectively as Gaussian plume solutions, have the distinct advantage of
being simple and relatively cheap to compute, and consequently form the
basis of many standard monitoring tools (including AERMOD and CALPUFF).
For more realistic situations where one is interested in incorporating
effects such as topographical variations or more complex time-varying
wind patterns, the only recourse is to solve the governing PDEs
directly, typically using approaches based on finite
volume~\cite{hosseini-mscthesis-2013, hosseini-FV-atmos}, finite
difference~\cite{lange-1978, nikmo-etal-1999} or finite element
schemes~\cite{albani-etal-2015}.

In comparison with the atmospheric dispersion models just described for
solving the forward problem, the source inversion problem has attracted
less attention in the literature.  The monograph by Vogel~\cite{vogel}
is a notable reference that lays out the general mathematical theory for
inverse problems as well as common solution approaches.  More specific
to the context of atmospheric transport, the work of Rao~\cite{rao-2007}
and Enting~\cite{enting-2002} provides an overview of methods for
solving the source inversion problem.  For example, inversion of
fugitive sources was considered recently in~\cite{sanfelix-etal-2015}
where the authors coupled a finite volume solver within an optimization
algorithm in order to estimate source emission rates.  Another powerful
and promising approach to the solution of inverse problems is based on
Bayesian methods, whose mathematical formulation is well-described by
Kaipio and Somersalo~\cite{somersalo} and
Stuart~\cite{stuart-acta-numerica}.  Although some attempts have been
made to apply Bayesian methods in the context of atmospheric dispersion
problems, their application is much less extensive than other
approaches.  Some examples include Huang et al.~\cite{Huang20151} who
employ a Bayesian method for identifying the location and emission rate
for a single point source by incorporating a Gaussian puff solution,
while Keats et al.~\cite{keats} use Bayesian inference to identify
emissions in a more complex urban environment. In both cases, the
authors use an adjoint approach to efficiently solve the
advection-diffusion PDE and evaluate the likelihood function using a
Markov Chain Monte Carlo algorithm. Senocak et al.~\cite{senocak} and
Wade and Senocak~\cite{wade} use Bayesian inference along with a
Gaussian plume model in order to reconstruct multiple sources in an
atmospheric contamination scenario.

In this paper, we aim to develop an accurate and efficient Bayesian
approach for solving the inverse source identification problem.  We aim
to estimate fugitive particulate emissions from various areas of an
industrial site based on measurements of contaminant concentration and
particulate mass deposited at a distance from the suspected sources.  We
consider a scenario wherein material from fugitive sources is dispersed
by the wind and then deposited on the ground due to a combination of
diffusive transport and gravitational settling.  Some sources may
reasonably be approximated as constant in time, but we are particularly
interested in the study of time-dependent sources arising for example
from dust entrained during loading operations that are performed on a
rotating shift schedule.  In order to monitor emissions in such a
scenario, various types of measurements are typically performed at fixed
locations in the vicinity of the known or suspected sources.  We are
particularly interested in two classes of measurements, deriving from
either total accumulated deposition of particles over a long time period
(on the order of one month) or short-time averaged concentrations
(taken over a period of one hour, which can be considered essentially
instantaneous in comparison with long-term measurements).  We utilize a
Gaussian plume model for short-range dispersion of pollutants and
incorporate this model within the Bayesian framework for solution of
inverse problems. We split the industrial site into a number of areas
that are suspected to contain the most important fugitive sources. The
Bayesian framework provides a natural setting for estimating emission
rates and also quantifying the uncertainty associated with the
estimates.  This study was motivated by a collaboration with Teck
Resources Ltd., in which we studied particulate emissions from a
lead-zinc smelter located in Trail, British Columbia,
Canada~\cite{hosseini-mscthesis-2013, stockie2010inverse}.

In contrast with some other studies, we do not consider the problem of
determining either the number or location of sources.  Instead, we
consider a given number of areas corresponding to suspected fugitive
emission sources, and approximate each area source by a single point
source located at the area centroid; however, we do allow the emission
rate for each source to vary as a continuous function of time.  We also
incorporate multiple measurement types and develop a unified framework
in which the forward model relates the entire measurement data set to
the emission rates.  The main challenge we encounter is in terms of the
low quality of the measured data that derives from two main features:
first, the most abundant measurements are from dust-fall jars, which
measure only monthly accumulated deposition and so unfortunately provide
no information about short-time variations in particulate emission; and
second, although we do have access to a few real-time measurement
devices, these sensors provide useful data only when the wind blows in
the direction linking them to the sources of interest. The main
advantage of our Bayesian framework is its ability to obtain a solution
even when the measured data is of relatively low quality and the problem
is severely under-determined.  Another benefit of our approach is that
we obtain an estimate of the accompanying uncertainty in the solution.
Finally, we stress that our framework can easily be extended to deal
with a general class of atmospheric dispersion problems including
applications such as seed or odour dispersal~\cite{bunton-etal-2007,
  levin-etal-2003}, natural disasters such as volcanic
eruptions~\cite{turner-hurst-2001}, and nuclear or chemical
accidents~\cite{miller-hively-1987}.

The remainder of this article is organized as follows. In
Section~\ref{sec:forward} we present the forward model which is based on
a Gaussian plume solution. In Section~\ref{sec:inverse} we develop the
Bayesian framework for solving the inverse problem by considering three
different instances of the inverse problem corresponding to different
models for the prior distribution. Section~\ref{sec:case-study} is
dedicated to an industrial case study in which we apply our framework
using a physical dataset in order to estimate sources of fugitive
emissions from a lead-zinc smelter.  After solving the inverse problem
we also assess the impact of the estimated sources by propagating the
solution of the Bayesian inverse problem through the forward model.

\section{The forward problem}
\label{sec:forward}

Dispersion and deposition of airborne particulates can be described
mathematically using the advection--diffusion equation 
\begin{equation}
  \label{advection-diffusion}
  c_t + \mb{u} \cdot \nabla c -\nabla \cdot ( \mb{K} \nabla c  )  = Q,
\end{equation}
which is a partial differential equation for the unknown particle
concentration, $c(\mb{x}, t) \: \mybunits{kg \:m^{-3}}$.  The other
quantities appearing in this equation are the wind advection velocity
$\mb{u} (\mb{x}, t) \: \mybunits{m \: s^{-1}}$, diffusion tensor
$\mb{K}(\mb{x}, t) := \diag( k_x, k_y, k_z) (\mb{x}, t) \:
\mybunits{m^2\: s^{-1}}$, and emission source term $Q(\mb{x}, t) \:
\mybunits{kg \: m^{-3} \: s^{-1}}$, which are all assumed to be given
functions.  We denote the spatial coordinates by $\mb{x} = (x, y, z)^T
\: \mybunits{m}$ and look for solutions of~\eqref{advection-diffusion}
defined on $\Omega \times [0,T]$, where the spatial variable
$\mb{x}\in\Omega = \{ -\infty \le x, y \le \infty, 0 \le z \le \infty
\}$ is in the half-space (assuming the ground is flat and located along
$z=0$), and the time interval of interest is $t\in[0,T]$.  We impose
$c(\mb{x},0)=0$ so that initially there is no contaminant within the
domain, and specify the consistent far-field boundary conditions
$c(\mb{x},t)\to 0$ as $|\mb{x}|\to \infty$.  At ground level we apply a
mixed (Robin) boundary condition for the deposition flux
\begin{equation}
  \label{robin-bc}
  \left. \left( K_z \frac{\partial c}{\partial z} + W_{\text{set}} c
    \right)   \right|_{z=0} =  W_{\text{dep}} c|_{z=0},
\end{equation}
which is commonly applied for capturing deposition of airborne
particulates onto the ground surface.  Here, $W_{\text{dep}} \: \mybunits{m \:
  s^{-1}}$ is the deposition velocity~\cite{seinfeld}, which is assumed
to be constant and dependent only on the individual particles under
consideration.  There is a corresponding settling velocity for each
particle type, $W_{\text{set}} \: \mybunits{m \: s^{-1}}$, which is also
constant and is approximated using the Stokes law for spherical
particles
\begin{equation}
  \label{stokeslaw}
  W_{\text{set}} = \frac{\rho g d_p^2}{18 \mu},
\end{equation}
where $\rho\: \mybunits{kg \: m^{-3}}$ is the mass density of the
particle, $d_p \:\mybunits{m}$ is particle diameter, $\mu = 1.8 \times 10^{-5} \:
\myunits{kg \: m^{-1} s^{-1}}$ is viscosity of air, and $g = 9.8
\:\myunits{m \:s^{-2}} $ is gravitational acceleration.  Since wind
measurements are typically only available at or near ground level, any
study of atmospheric particulate transport requires choosing some
appropriate model for the wind velocity $\mb{u}$, such as a power-law or
logarithmic profile for the vertical variation~\cite{arya,
  seinfeld}. Likewise, many models are available to represent the
diffusion coefficients $\mb{K}$ as functions of height and other
solution variables, such as the Monin-Obukhov model~\cite{seinfeld}.

Once all parameters are specified, a common approach to
solving \eqref{advection-diffusion} is by discretizing $\Omega$ on a
finite computational domain, imposing some suitable artificial boundary
conditions, and then applying a numerical method that is capable of
handling variable-coefficient parabolic partial differential equations.
Examples of commonly-used algorithms for atmospheric dispersion models
include finite volume, finite element and spectral
methods~\cite{albani-etal-2015, christensen-prahm-1976,
  hosseini-FV-atmos, lee1997computational, mcrae1982numerical}.
% leveque,
However, in this study our interest is in solving inverse problems,
which requires many evaluations of the solution of
\eqref{advection-diffusion} and hence a direct application of any such
algorithm is often too expensive.  Instead, we will make use of an
approximate analytical solution of~\eqref{advection-diffusion} known as
the \emph{Gaussian plume solution}.  Methods based on the Gaussian plume
approximation are widespread in the literature, as well as in both
commercial and public domain software used for regulatory and monitoring
purposes~\cite{arya-1995, hanna-briggs-hosker-1982, seinfeld, epa-2010}.

\subsection{Gaussian plume solution}

We begin by imposing several simplifying assumptions on the problem that
allow us to use a large class of Gaussian plume solutions:

\begin{assumption-non}
  % \label{setup-assumption}
  \begin{enumerate}[{A}1.] 
  \item Wind velocity takes the form $\mb{u}(\mb{x},t) = (u_x(\mb{x},t),
    \; u_y(\mb{x},t), \; -W_{\text{set}})$ so that aside from a
    small vertical settling component, the wind blows only in the
    horizontal direction.
    \label{setup-assumption-a}
  \item Variations in topography are negligible, so that the ground may
    be treated as a horizontal plane.
    \label{setup-assumption-b}
  \item Only dry deposition occurs, and we neglect any
    enhancement in deposition due to rainfall or other events.
    \label{setup-assumption-c}
  \item The transient behavior of contaminant plumes can be neglected,
    and the solution is assumed to be in a quasi-steady state during
    each time interval in which wind measurements are taken.
    \label{setup-assumption-d}
  \item The site of interest can be divided into a number of disjoint
    area sources, and the emissions within each area are
    well-approximated by a single point source located at the area
    centroid.
    \label{setup-assumption-e}
  \end{enumerate}  
\end{assumption-non}

The first three Assumptions
A\ref{setup-assumption-a}--A\ref{setup-assumption-c} are imposed due to
limits on the availability of data.  For example, wind data are often
measured at only a few meteorological stations, and typically only the
horizontal wind components are available.  Variations in
topography are often difficult to measure accurately and, even more
importantly, it is difficult (or impossible) to reconstruct a wind field
over variable topography when wind measurements are sparse.  Finally,
the theory of wet deposition introduces significant complications and
requires meteorological data that are simply not available in many
applications.  Assumption A\ref{setup-assumption-d} is imposed so that a
Gaussian plume approximation for \eqref{advection-diffusion} can be
employed within each time step of a simulation.  In effect, the Gaussian
plume solution is the Green's function of the steady state
advection-diffusion equation for a specific choice of wind profile and
diffusion tensor~\cite{seinfeld, stockie2011siam}.  Assumption
A\ref{setup-assumption-e} is imposed to allow treatment of a general
class of sources, including well-defined areas of the site over which
emissions are distributed (such as large debris piles), and also sources
whose location cannot be accurately determined (such as large buildings
having many open vents, windows, etc.).
% Although this last
%assumption may seem particularly questionable, we show in
%\ref{appendixA} that it can be justified in settings where the
%measurements of concentration are taken significantly far away from the
%emission sources.  
Assumption A\ref{setup-assumption-e} also permits us to write the source
term $Q$ as a linear superposition of singular point sources
\begin{equation}
  \label{point-sources}
  Q(\mb{x}, t) = \sum_{i=1}^{N_s} q_i(t) \, \delta(\mb{x} - \mb{x}_i),
\end{equation}
where $q_i(t)$ is the (possibly time-varying) total emission rate from
the $i$th area, $\mb{x}_i$ is the location of the area centroid, $N_s$
is the total number of sites, and $\delta(\mb{x})$ represents the
three-dimensional delta distribution.  Owing to the linearity of both
Equation~\eqref{advection-diffusion} and the source term, we may
consider separately the contribution to the concentration field
$c_i(\mb{x},t)$ arising from each individual point source $i$, and then
write $c=\sum_{i=1}^{N_s} c_i$.

We also consider a scenario where wind measurements are available at a
sequence of times $t_j$, and each interval $[t_j, t_{j+1}]$ is chosen so
that Assumption~A\ref{setup-assumption-d} is satisfied; in practice,
taking a constant time interval of 10~minutes is reasonable~\cite{hanna-briggs-hosker-1982, stockie2010inverse}.  We then
introduce transformed spatial coordinates $\tilde{\mb{x}}_{i,j} =
(\tilde{x}, \tilde{y}, \tilde{z})$ with
\begin{equation}
  \label{coordinate-rotation}
  \tilde{\mb{x}}_{i,j} = \mb{R}_j ( \mb{x} - \mb{x}_i), \quad \mb{R}_j = 
  \begin{bmatrix}
    \cos(\theta_j) & -\sin(\theta_j)\;\;\; & 0\; \\
    \sin(\theta_j) & \cos(\theta_j) & 0\; \\
     0 & 0 & 1\;
  \end{bmatrix},
\end{equation}
and $\theta_j = \tan^{-1} (u_y(t_j)/u_x(t_j))$.  Note that
$\tilde{\mb{x}}_{i,j}$ represents a rotation of $\mb{x}$ so that the
$\tilde{x}$-axis points in direction of the horizontal wind vector at
time $t_j$, composed with a translation that shifts the $i$th source to
the origin. After this transformation, we may apply the Gaussian plume
solution of Ermak~\cite{ermak1977analytical, stockie2011siam} for a
point source located at the origin with wind speed $U=\left(u_x(t_j)^2 +
  u_y(t_j)^2\right)^{1/2}$ directed along the $\tilde{x}$-axis to obtain
\begin{equation}
  \label{ermak-solution}
  \begin{aligned} 
    & c_i(\tilde{\mb{x}}_{i,j}, t) =  \frac{q_i(t_j)}{2 \pi U
      \sigma_{\tilde{y}} \sigma_{\tilde{z}}}  \;
    \exp \left( -\frac{\tilde{y}^2}{2 \sigma_{\tilde{y}}^2}
      -\frac{W_{\text{set}} \tilde{z}}{2 K_{\tilde{z}}} -
      \frac{W_{\text{set}}^2 \sigma_{\tilde{z}}^2}{8 K_{\tilde{z}}^2}
    \right)  \\ 
    & \quad\times \left[ 
      \exp \left( -\frac{\tilde{z}^2}{2 \sigma_{\tilde{z}}^2} \right)  
      + \exp \left( -\frac{(\tilde{z} + 2z_i)^2}{2 \sigma_{\tilde{z}}^2} \right) 
      -\exp \left( \frac{W_o(\tilde{z} + 2z_i)}{ K_{\tilde{z}}} + \frac{W_o^2
          \sigma_{\tilde{z}}^2}{2 K_{\tilde{z}}^2}\right) \text{erf} \left(
        \frac{W_o\sigma_z}{ \sqrt{2}K_{\tilde{z}}} + \frac{\tilde{z} +
          2z_i}{\sqrt{2} \sigma_{\tilde{z}}^2}\right)  \right]. 
  \end{aligned}
\end{equation}
Here, $W_o := W_{\text{dep}} - \frac{1}{2} W_{\text{set}}$, the time
interval is $t \in [t_j, t_{j+1}]$, and the standard deviations of
concentration are given by
\begin{equation}
  \label{standard-deviations}
  \sigma^2_{\tilde{y},\tilde{z}}(\tilde{x}) = \frac{2}{U}
  \int_0^{\tilde{x} } K_{\tilde{y},\tilde{z}}(s)\; ds, 
\end{equation}
in terms of the diffusion coefficients $K_{\tilde{y}, \tilde{z}}$ in the
$\tilde{\mb{x}}$ coordinates. Note that the standard deviations
$\sigma_{\tilde{y},\tilde{z}}$ are assumed to be functions of downwind
distance $\tilde{x}$, and in practice are much easier to measure than
$K_{\tilde{y}, \tilde{z}}$.  Indeed, many formulas have been proposed in
the literature for $\sigma_{\tilde{y},\tilde{z}}$ based on theory and
experiment, and we will use one of the more common parameterizations due
to Briggs~\cite{briggs} with
\begin{equation}
  \label{Briggs-sigma}
  \sigma(\tilde{x}) = a  \tilde{x} ( 1 + b \tilde{x})^{-c}.
\end{equation}
Values of the parameters $a$, $b$ and $c$ are listed in
Table~\ref{tab:briggs-coeff} for different Pasquill atmospheric
stability classes.
\begin{table}[htp]
  \centering
  \caption{Value of parameters in \eqref{Briggs-sigma} for different
    Pasquill atmospheric stability classes~\cite{seinfeld}.} 
  \label{tab:briggs-coeff} 
  \begin{tabular}[htp]{c|ccc|ccc}
    \hline
    Stability class & \multicolumn{3}{|c|}{$\sigma_{\tilde{y}}$}
    & \multicolumn{3}{|c}{$\sigma_{\tilde{z}}$} \\ \cline{2-7}
    &   $a$    & $b$  & $c$  & $a$   & $b$    & $c$ \\ \hline
    A & 0.22 & 1.0e-4 & 0.50 & 0.20  & 0.0    & 0.0 \\
    B & 1.60 & 1.0e-4 & 0.50 & 1.2   & 0.0    & 0.0 \\
    C & 0.11 & 1.0e-4 & 0.50 & 0.08  & 2.0e-4 & 0.5 \\
    D & 0.08 & 1.0e-4 & 0.50 & 0.06  & 1.5e-3 & 0.5 \\
    E & 0.06 & 1.0e-4 & 0.50 & 0.03  & 3.0e-4 & 1.0 \\
    F & 0.04 & 1.0e-4 & 0.50 & 0.016 & 3.0e-4 & 1.0 \\ \hline
  \end{tabular}
\end{table}

Note that equation \eqref{ermak-solution} remains linear in the
emission rates $q_i(t)$ so that we can simply rotate the Ermak solution
back into the original $\mb{x}$-coordinate system to obtain
\begin{equation}
  \label{forward-linear-form}
  c( \mb{x}, t) = \sum_{i=1}^{N_s} q_i(t_j) \, \mathcal{G}_i( \mb{x} ;\;
  \mb{u}(\mb{x},t_j) , \mb{x}_i, W_{\text{set}}, W_{\text{dep}},
  \sigma_{\tilde{x}}, \sigma_{\tilde{y}} ), \quad t \in [t_j, t_{j+1}),  
\end{equation}
for suitably defined functions $\mathcal{G}_i$.  From this point on,
we will suppress the dependence of $\mathcal{G}_i$ on all parameters
other than $\mb{x}$ in order to obtain a cleaner notation. This is also
convenient because for a given set of wind data and other parameters,
the $\mathcal{G}_i$ are independent of $q_i(t_j)$.

Equation~\eqref{forward-linear-form} concludes our approximation of the
solution to the forward problem, which we use to compute the
concentration at any given time based on knowledge of emission rates and
other parameters.  We have so far assumed that $t_j$ are the times at
which the wind velocity is measured.  However, we can easily generalize
to the situation where wind, concentration and emission rates are at
different times by making use of either interpolation or averaging. We
will return to this issue in Section~\ref{sec:case-study} and construct
an arbitrary time mesh by first fitting a Gaussian
process~\cite{rasmussen} to the wind data and then evaluating the mean
of this process on a discrete time grid.

\subsection{Modelling observations}

We next describe how to incorporate various types of
measurements of deposition or concentration that are commonly encountered in practical applications,
aiming for a general framework that is capable of handling both
short-time (instantaneous) and long-time (accumulated or integrated)
sensor measurements.  For this purpose, we construct a linear operator
that represents the transformation between unknown emission sources
$q_i(t_j)$ and known deposition measurements.

Suppose that a total of $N_r$ sensors are located throughout the domain
in the vicinity of the sources of interest, and let $\mb{\bar{x}}_k$ for
$k = 1, \dots, N_r$ represent the sensor locations.  We let $d_{\ell,k}$
represent the corresponding measurement
% (of deposition or concentration)
at $\mb{\bar{x}}_k$, where the subscript $\ell=1,\cdots, m_k$ refers to
data from the $k$-th sensor corresponding to the $\ell$-th time interval.
% We assume that the $k$th sensor can
% be modelled as a linear functional of the form $\mathcal{M}_k =
% \delta(\mb{x} - \mb{\bar{x}}_k) \mathcal{T}_k(t)$ where
% $\mathcal{T}_k(t)$ is an appropriate functional that only depends on
% time and will be identified
% for each sensor below.
% The functional $\mathcal{M}_k$ acts on the concentration $c(\mb{x},t)$
% for $(\mb{x}, t) \in \Omega \times (0,T]$ and returns a real vector of
% length $m_k$; that is, $\Range (\mathcal{M}_k) = \reals^{m_k}$, where
% $m_k>1$ for time-dependent measurements and $m_k=1$ otherwise.  
% Note that the Ermak solution \eqref{ermak-solution} is infinitely smooth
% away from the origin, which translates into the concentration being
% infinitely smooth away from the source locations in
% \eqref{forward-linear-form}.  Therefore, applying $\delta(\mb{x} -
% \mb{\bar{x}}_k)$ to $c$ is well-defined as long as the sensor positions
% do not coincide with any of the sources; that is, $\mb{\bar{x}}_k \in
% \Omega \setminus \cup_{i=1}^{N_s}\{\mb{x}_i \}$. As long as this
% assumption holds,
%In practice, even time-dependent measurements are averages over some
%short time interval, and so
 We assume that each measurement can be represented as a time integral
\begin{equation}
  \label{data-kernel-form}
  d_{\ell,k} = \int_0^T f_{\ell,k}(t) \, c( \mb{\bar{x}}_k, t)\; dt,
\end{equation}
where the $f_{\ell,k}(t)$ are ``window functions'' that pick out the
corresponding time interval during which each sensor actively
measures the concentration or deposition. 
%Then the functional $\mathcal{T}_k$ is defined via
%\eqref{data-kernel-form} and the choice of the kernels $f_{\ell,k}$.
A different kernel $f_{\ell,k}$ must be chosen for each measurement
device, and these functions are discussed in the next two subsections
for the dust-fall jars and real-time measurement devices used in this
study.  In practice, one needs to approximate the above integrals
numerically, for which we propose using a time discretization based on a
uniform grid $\{t_i\}_{i=1}^{N_T}$ that coincides with the time
intervals used for computing concentration in
\eqref{forward-linear-form}.  We then employ a simple one-sided
quadrature rule to approximate \eqref{data-kernel-form}.

Collecting all sensor measurements into a vector
\begin{equation}
  \label{measurement-map}
  \mb{d}_k  := (d_{1,k}, d_{2,k}, \dots, d_{m_k, k})\in
  \reals^{m_k},  
\end{equation}
we then write a matrix representation that connects
concentrations to sensor measurements:
\begin{equation}
  \label{linear-measurement-1}
  \mb{d}_k = \mb{M}_k \mb{c}_k \quad \text{where} \quad 
  \mb{c}_k := (c(\bar{\mb{x}}_k, t_1), \dots, c(\bar{\mb{x}}_k, t_{N_T}))^T.
\end{equation}
%\todo{[Are the $\mb{c}_k$ values at $\mb{x}_k$ or $\mb{\bar{x}}_k$?]}
Exploiting the linearity of \eqref{forward-linear-form}, we may then
define a block diagonal matrix form of the functions $\mathcal{G}_i$
\begin{equation}
  \label{vectors-intermediate}
  \mb{G}_k := \diag( \mb{g}_{k,1} , \mb{g}_{k,2}, \dots,
  \mb{g}_{k,N_T} ) \quad \text{where} 
  \quad 
  \mb{g}_{k,j} := (\mathcal{G}_1(\mb{\bar{x}}_k; t_j), \dots,
  \mathcal{G}_{N_s}(\mb{\bar{x}}_k; t_j)),  
\end{equation}
which in turn allows us to write
\begin{equation}
  \label{linear-measurement-single}
  \mb{d}_k = \mb{M}_k \mb{G}_k \mb{q} \quad\text{where}\quad \mb{q} := ( q_1(t_1),
  \dots, q_{N_s}(t_1), q_1(t_2), \dots, q_{N_s}(t_2), \dots, 
  q_1(t_{N_T}), \dots, q_{N_s}(t_{N_T}) )^T. 
\end{equation}
We may then concatenate the $\mb{d}_k$ vectors for all measurement
devices, and similarly for the $\mb{M}_k$ and $\mb{G}_k$ matrices, which
allows us to rewrite \eqref{linear-measurement-single} in the compact
form
\begin{equation}
  \label{linear-measurement}
  \mb{d} = \mb{M} \mb{G}\mb{q} =: \mb{F}\mb{q},
\end{equation}
where $\mb{d}$ is a long vector containing all available data from
measurement devices, $\mb{q}$ is the discretization of $q_i(t)$
functions in time, and we refer to $\mb{F} = \mb{M} \mb{G}$ as the {\it
  observation map} or {\it matrix}. In the next two sections, we define
the window functions $f_{\ell,k}$ appearing in \eqref{data-kernel-form}
for the two specific types of measurement devices used in this study.

\subsubsection{Dust-fall jars} 

Dust-fall jars provide a cheap and convenient means of measuring
long-term deposition of solid particulate matter such as the lead and
zinc oxides that are of particular concern here.  These jars are simply
cylindrical plastic containers that are filled with water and left out
in the open, usually on an elevated platform, for an extended time
period (in our case, for one month). Sometimes a mesh is placed on top
of the jar to prevent large pieces of unwanted material from
contaminating the sample (e.g., leaves, garbage or insects).  At the end
of a measurement period, the jars are sent to a laboratory for chemical
analysis, yielding a list of particles of interest and their total mass.
Each dust-fall jar provides a single data point for every sampling
period, which corresponds to taking $\ell=\{1\}$ in
\eqref{data-kernel-form}.  Suppose that each jar has cross-sectional
area $A \: \mybunits{m^2}$ and recall from \eqref{robin-bc} that the
ground-level deposition flux is given by $W_{\text{dep}} c(\mb{x},t)$.
Then a dust-fall jar can be treated within the general framework 
developed above by taking $f_{\ell,k} = A W_{\text{dep}} \mb{1}_{(0,T]}$
where $\mb{1}_{(0,T]}$ is the indicator function for the entire time
interval of interest.

\subsubsection{Real-time measurement devices}

The other major class of measurement devices are real-time sensors such
as the Xact ambient metals monitor~\cite{cooper-environmental-2015}
or Andersen high-volume air sampler~\cite{thermo-scientific-2015}
that measure average concentration of particulates over a shorter time
period ranging from minutes to hours.  These devices are often operated
automatically according to a preset schedule, coming on-line and taking
measurements at a set of specified times, and otherwise remaining
inactive.  Suppose that such a sensor is scheduled to initiate
measurements at a sequence of times $\{\tau_1, \tau_2, \dots, \tau_p\}$
and that each measurement of average concentration is taken over a time
period of length $\Delta \tau$. Then we can write $f_{\ell,k} =
\frac{1}{\Delta \tau} \mb{1}_{(\tau_\ell,\; \tau_\ell+ \Delta \tau]}$ with
$\ell = \{ 1, 2, \dots, p \}$.

\section{The inverse problem}
\label{sec:inverse}

In the previous section, we considered the forward problem and derived
an explicit form for the linear mapping that takes as input the emission
rates from all sources and returns as output the ground-level deposition
at a number of specified sensor locations.  Our actual aim is to solve
the inverse problem, which is equivalent to inverting this linear
mapping and corresponds to taking a set of sensor observations at given
locations and estimating the corresponding emission rates.  However,
before getting into the details we first need to address the issue of
noise due to measurement errors, which can have a major impact on the
solution to the inverse problem.

%\subsection{Measurement noise}
%\label{sec:noise}

Assuming that real measurements $\tilde{\mb{d}}$ are actually noisy
realizations of the predictions of our model, we need to replace 
Equation~\eqref{linear-measurement} with 
\begin{equation}
  \label{noisy-forward-prob}
  \mb{d}   = \mb{F} \mb{q} + \pmb{{\epsilon}},
\end{equation}
where $\pmb{\epsilon}$ is a vector of independent, additive
noise. Throughout this paper we will assume that the noise is a
zero-mean multivariate Gaussian random variable, $\pmb{{\epsilon}} \sim
N(0, \pmb{\Sigma})$.  Note that the covariance operator $\pmb{\Sigma}$
is taken to be diagonal owing to the independence assumption; however,
the diagonal elements may still vary between data points owing to
differences in the accuracy of each sensor. 

Our approach to solving the source inversion problem with additive noise
will proceed in three steps, corresponding to three problems that
impose progressively more realistic constraints on the solution
and hence become increasingly more challenging.  The main reason for
using this approach is that the data available for the problem of
interest is actually quite sparse (which we describe in
Section~\ref{sec:case-study}).  Therefore, the prior knowledge of the
solution (modelled via the prior distribution defined below) has a
strong influence on the final solution.  In order to come up with a good
prior distribution, we first employ the maximum likelihood (MLE) or
posterior mean estimate for simpler problems in order to construct a
prior distribution for the next, more complex, problem.

\subsection{Constant emissions}
\label{sec:constant-emission} 

Our first step in solving the source inversion problem assumes that
emission rates are constant in time.  In order to apply the framework 
outlined above, it is convenient to define a vector
$\mb{\tilde{p}}_{\text{c}}$ of length $N_s$ containing the list of
constant emission rates and then introduce a matrix $\mb{A}$ such that
$\mb{\tilde{q}}_{\text{c}} := \mb{A} \mb{\tilde{p}}_{\text{c}}$, where
$\mb{A}$ contains $N_T$ copies of each entry in the emissions vector
$\mb{\tilde{p}}_c$ so that $\mb{\tilde{q}}_c$ corresponds to a
straightforward discretization of the constant-in-time emissions problem
(hence the subscript $c$).

We assumed that the noise $\pmb{\epsilon}$ is a multivariate zero-mean
Gaussian random variable that is distributed according to the density
\begin{equation}
  \label{gaussian-density}
  \pi_{\pmb{\epsilon}} (\mb{v}) = \left((2\pi)^M
    |\pmb{\Sigma}|\right)^{-1/2} 
    \exp\left( -\frac{1}{2} \| \Sigma^{-1/2} \mb{x} \|_2^2 \right),
    \qquad \mb{v} \in \reals^M.
\end{equation}
Combining this with \eqref{noisy-forward-prob} gives
\begin{equation}
  \label{constant-emission-likelihood}
 \pi(\mb{d} \,|\, \mb{\tilde{q}}_{\text{c}} ) 
= \pi(\mb{d} \,|\, \mb{\tilde{p}}_{\text{c}}) =
\pi_{\pmb{\epsilon}}(\mb{d}- 
  \mb{F}\mb{A} \tilde{\mb{p}}_c) = \left((2 \pi)^M
  |\pmb{\Sigma}|\right)^{-1/2} \exp \left(
    -\frac{1}{2}  \left\|    \pmb{\Sigma}^{-1/2} (\mb{F} \mb{A}
      \mb{\tilde{p}}_{\text{c}} - 
      \mb{d}) \right\|_2^2 \right),
\end{equation}
where $M$ denotes the length of the vector $\mb{d}$,
$\pmb{\Sigma}^{1/2}$ is defined in the sense of the square root of
non-negative definite matrices, and $\|\cdot\|_2$ is the usual Euclidean
norm.
%\todo{[WHOA!  This came out of the blue.  The
 % ``likelihood distribution'' needs some extra introduction.]}  
Given that the parameter space in this setting is small (with size equal
to the number of point sources) it is sufficient to use a maximum
likelihood estimator. We therefore consider the maximizer of the
distribution on the right hand side of
\eqref{constant-emission-likelihood} which corresponds to the solution
of the following constrained least squares problem:
\begin{equation}
  \label{MLE-estimate}
  \mb{q}_{\text{c}} := \mb{A} \mb{p}_{\text{c}}, \qquad
  \mb{p}_{\text{c}} := \argmin_{\mb{\tilde{p}}_{\text{c}}}  \left\|
    \pmb{\Sigma}^{-1/2} (\mb{F} \mb{A} \mb{\tilde{p}}_{\text{c}} - 
    \mb{d}) \right\|_2^2 \quad \text{subject to }
  {\mb{\tilde{p}}_{\text{c}} \ge 0}. 
\end{equation}
This problem can be solved efficiently using standard software
such as MATLAB's lsqlin function (which we use here).

\subsection{Unconstrained emission rates}
\label{sec:unconstrained}

In the second step, we employ the solution just derived for the constant
emissions case to construct a prior distribution for the less
restrictive problem in which the emission rates are not constant but
rather vary smoothly in time (we leave the non-negativity constraint for
the third and last step).  This results in a linear problem for which we
can still write an explicit analytical solution.  We first rewrite
the likelihood distribution appearing in
\eqref{constant-emission-likelihood} as
\begin{equation}
  \label{likelihood}
  \pi\left( \mb{{d}} \,|\, \mb{\tilde{q}}_{\text{s} }\right) = 
  \left((2 \pi\right)^M
  |\pmb{\Sigma}|)^{-1/2} \exp\left( -\frac{1}{2} \left\|
      \pmb{\Sigma}^{-1/2} (\mb{F}\mb{\tilde{q}}_{\text{s}} - \mb{d})
    \right\|_2^2 \right). 
\end{equation}
Our intention is to use Bayes' rule~\cite{somersalo} to estimate the
variable emission rates $\mb{q}_{\text{s}}$ given the sensor data
$\mb{d}$, where the subscript ``s'' stands for ``smoothness''.  We then
use the MLE solution from \eqref{MLE-estimate} to construct a Gaussian
prior distribution on $\mb{q}_{\text{s}}$ as
\begin{equation}
  \label{prior}
  \pi_{\text{prior}} (\mb{\tilde{q}}_{\text{s}}) = 
  \left( (2 \pi\right)^N
  |\mb{C}|)^{-1/2}\exp\left( -\frac{1}{2}  
    \| \mb{C}^{-1/2} (\mb{\tilde{q}}_{\text{s}} -  
    \mb{q}_{\text{c}}) \|_2^2 \right),
\end{equation}
where the choice of the covariance matrix $\mb{C}$ is the key to
constructing an appropriate prior.  Here, we assume that $\mb{C}$
has a block diagonal structure
\begin{equation}
  \label{covariance-form}
  \mb{C} = \mb{I}_{N_s} \otimes \mb{L}^{-2},
\end{equation}
where $\mb{I}_{N_s}$ is the identity matrix of size $N_s\times N_s$
(recall $N_s$ is the number of sources), $\otimes$ denotes the Kronecker
product, and $\mb{L}$ is a finite difference discretization of the
differential operator
\begin{equation}
  \label{differential-operator}
  \mathcal{L} := \alpha ( \mathcal{I} - \gamma \partial_{tt} ) \quad
  \text{on} \quad [0,T].
\end{equation}
Here, $\alpha, \gamma > 0$ are fixed constants, $\mathcal{I}$ is the
identity map, and $\partial_{tt}$ is the Laplacian with homogeneous
Neumann boundary conditions. Intuitively, this choice implies that the
second derivative of the emission rates is distributed as a Gaussian
random variable with bounded variance at each point and so it yields
sufficiently smooth sample functions.

\begin{figure}[htp]
  \centering
  \begin{tabular}{c@{}c}
    \includegraphics[height=0.2\textheight]{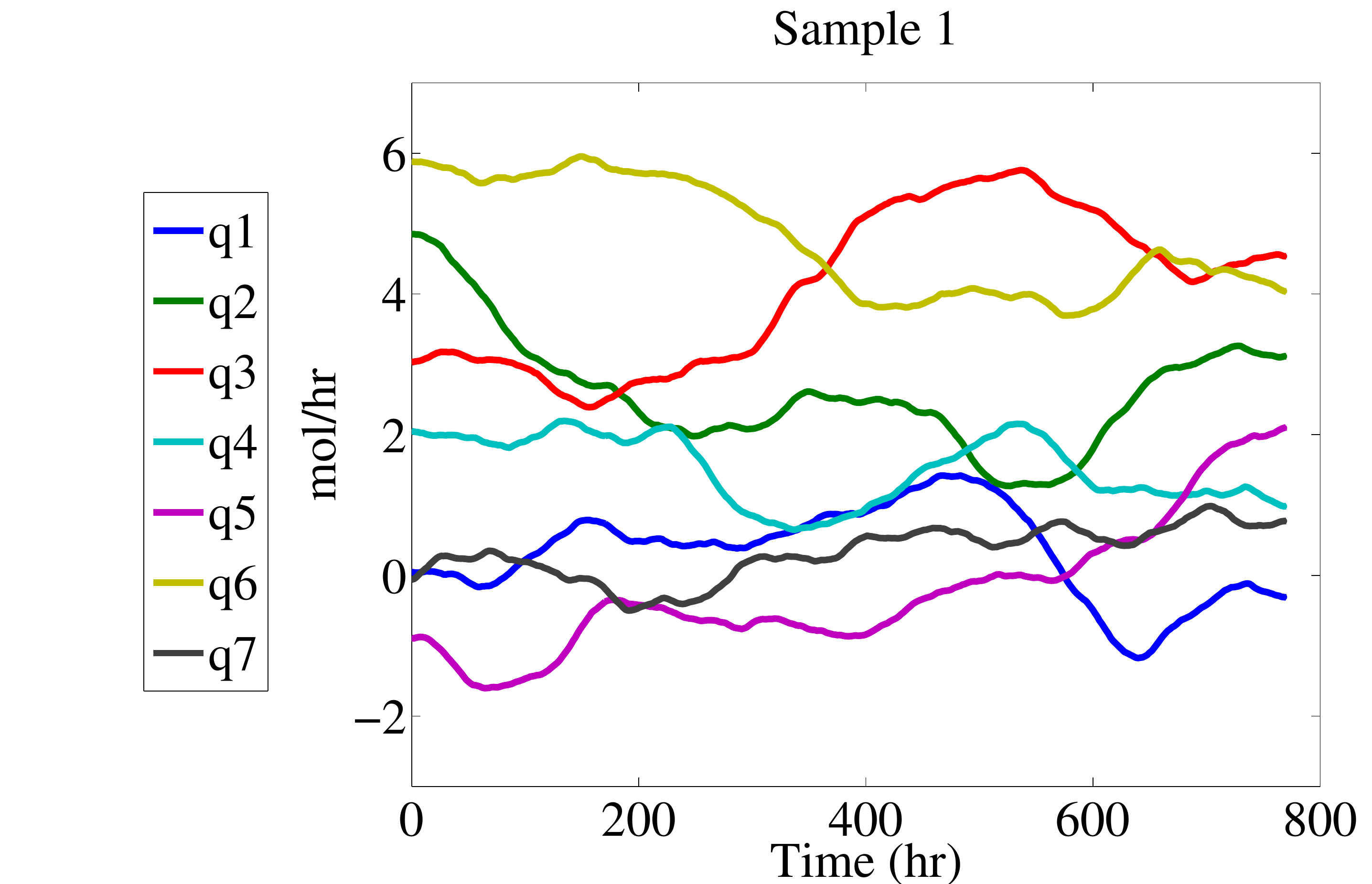}
    & 
    \includegraphics[height=0.2\textheight]{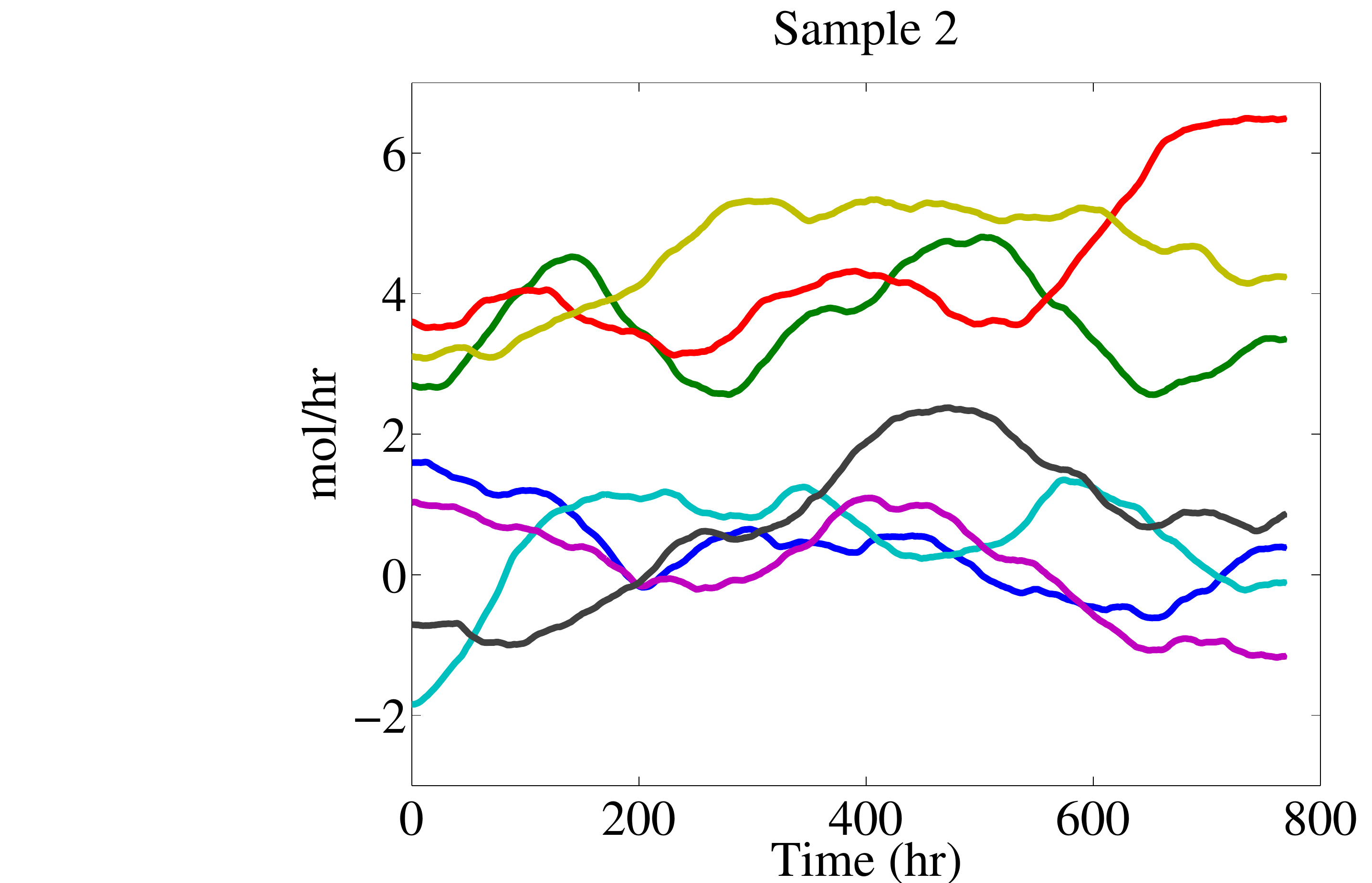}
  \end{tabular}
  \caption{Two samples from the zero-mean distribution $N(\mb{0},\,
    \mb{I}_{N_s} \otimes \mb{L}^{-2})$ for the synthetic data problem,
    with $\alpha = 1$ and $\gamma = 5\times10^{-3}$.}
  \label{fig:samples}
\end{figure}

It remains to choose a suitable discretization of the operator
$\mathcal{L}$ in \eqref{differential-operator}, for which we use finite
differences to get
\begin{equation}
  \label{discretized-covariance}
  \mb{L} = {\alpha}{\sqrt{\frac{\Delta t}{T}}} ( \mb{I} - \gamma \pmb{\Delta})
  \qquad \text{and} \qquad
  \pmb{\Delta} := \left( \frac{T}{\Delta t} \right)^2
  \begin{bmatrix}
    -1 & 1 & 0 & & \cdots \\
    1 & -2 & 1 & 0 & \cdots \\
    & \ddots & \ddots & \ddots & \\
    \cdots  & 0 &  1 & -2  &1\\
    \cdots &  0& 0& 1& -1 
  \end{bmatrix}.
\end{equation}
The parameter $\alpha$ controls the prior variance whereas $\gamma$
controls the bandwidth of samples in the Fourier domain. Our choice of
scaling $\sqrt{\Delta t/T}$ ensures that we obtain a proper
discretization of the biharmonic operator, in which sample variances
are independent of the discretization parameter $\Delta t$\
%\begin{gather*}
 % \mb{u} ^T\mb{L}^2 \mb{u} \to \left(\|u(\tau)\|^2_{L^2([0,1])} +
  %  \|\partial_\tau u (\tau)\|^2_{L^2([0,1])} + \|\partial_{\tau \tau} u
   % (\tau)\|^2_{L^2([0,1])} \right) \quad \text{as} \quad \Delta \tau
  %\to 0,
%\end{gather*}
%where $\tau := t/T$, $u(t)$ is any given function on $[0,T]$, and
%$\mb{u}$ is a finite difference discretization of $u$.  Here,
%$\|u\|_{L^2([0,1])} := (\int_0^1 |u(\tau)|^2 d\tau)^{1/2}$ \todo{[Should
  %this norm be on $L^2([0,T])$?  Or else re-define $u$ as a function of
 % $\tau$.]} is the usual $L^2$-norm.  The scaling $\tau = t/T$ is
%introduced to normalize the first eigenvalue of the operator so that the
%shape of the samples does not depend on the length of the time interval
(for more details on discretization of random functions,
see~\cite[Sec.~5.7]{somersalo} and \cite{lassas}).  Throughout the rest
of this paper, we choose $\alpha =1$ and $\gamma=5 \times 10^{-3}$ for the two
control parameters in \eqref{discretized-covariance}.
Figure~\ref{fig:samples} depicts two samples from the zero-mean prior
distribution for this parameter choice, and we draw the reader's
attention to both the continuity and smoothness of the samples.
%\todo{ Rewrite this section
 % (perhaps with more details) so that it is accessible to
 % atmos. env. audience.}

%This choice of the prior covariance
%implies that draws from \eqref{prior} are independent
%$2p$-differentiable functions in time (refer
%to~\cite[Sec.~3.3]{somersalo} for more information on regularity of
%samples and construction of priors).

With the prior distribution identified, we can now use Bayes' rule to
write the posterior distribution of $\mb{\tilde{q}}_{\text{s}}$ given
the data as
\begin{equation}
  \label{Bayesrule-posterior}
  \pi_{\text{posterior}} (\mb{\tilde{q}}_{\text{s}} \,|\, \mb{d} )
  \propto \exp\left( -\frac{1}{2} \left\| \pmb{\Sigma}^{-1/2}
      (\mb{F}\mb{\tilde{q}}_{\text{s}} - \mb{d}) \right\|_2^2
  \right) \exp\left( -\frac{1}{2}  \left\| \mb{C}^{-1/2}
      (\mb{\tilde{q}}_{\text{s}}-\mb{q}_{\text{c}}) 
    \right\|_2^2 \right).
\end{equation}
Because \eqref{forward-linear-form} is linear in $\mb{\tilde{q}}_s$, 
both the likelihood and the prior are Gaussian and so the posterior
must also be Gaussian.  In this setting, we may write the posterior
analytically~\cite{rasmussen, somersalo} as
\begin{equation}
  \label{Gaussian-post}
  \pi_{\text{posterior}} (\mb{\tilde{q}}_{\text{s}} \,|\, \mb{d} ) =
  N(\mb{q}_{\text{s}}, \mb{C}_{\text{s}}) ,
\end{equation}
where 
\begin{subequations} \label{posterior-mean-cov}
  \begin{align}
   \mb{q}_{\text{s}} & := \mb{q}_{\text{c}} + \mb{C} \mb{F}^T
   (\pmb{\Sigma} + \mb{F}
    \mb{C} \mb{F}^T)^{-1} (\mb{d} - \mb{F} \mb{q}_{\text{c}}),
    \\
  % \mb{q}_{\text{s}} & := \mb{q}_{\text{c}} + (\mb{d} - \mb{F}
  %                     \mb{q}_{\text{c}})     (\pmb{\Sigma} + \mb{F}
 %    \mb{C} \mb{F}^T )^{-T}
% \mb{C} \mb{F}^T 
%,\\
    \mb{C}_{\text{s}} & := \mb{C} - \mb{C} \mb{F}^T (\pmb{\Sigma} +
    \mb{F} \mb{C} \mb{F}^T )^{-1} \mb{F} \mb{C},
  \end{align}
\end{subequations}
which completely determines the posterior distribution for the emission
rates. Here, we take $\mb{q}_{\text{s}}$ to be our next best guess of
the true emission rates, which is intuitive because the mean of a
Gaussian distribution coincides with the maximizer of its distribution
so that $\mb{q}_{\text{s}}$ is also the point of maximum probability of
the posterior distribution (this time with the smoothness prior). We can
also think of $\mb{q}_s$ as an improvement on the constant-emissions
estimate $\mb{q}_c$, which permits the emission rates to vary in time but
still aims to keep the average emission rates close to $\mb{q}_c$.

\subsection{Non-negative emission rates} 
\label{sec:non-negative}

In the third and final step, we impose a positivity (or more accurately
non-negativity) constraint on the emission rates, for which the forward
problem now becomes nonlinear.  We define the function
\begin{equation}
  \label{source-level-set}
  h: \reals^N \to \reals^N, \quad  \mb{q} =  {h}( \mb{v} ) \quad
  \text{then} \quad \mb{q}_i = \max\{ 0, \mb{v}_i\}, 
\end{equation}
which permits us to pose the inverse problem for an auxiliary vector
$\mb{v}$ and rewrite \eqref{Bayesrule-posterior} as
\begin{equation}
  \label{Bayesrule-posterior-positive}
  \pi_{\text{posterior}} (\mb{v} \,|\, \mb{d} ) \propto \exp\left(
    -\frac{1}{2} \left\| \pmb{\Sigma}^{-1/2} (\mb{F}h(\mb{v)} -
      \mb{d}) \right\|_2^2 \right) \exp\left( -\frac{1}{2}  \|
    \mb{C}^{-1/2} (\mb{v}-
    h(\mb{q}_{\text{s}}) \|_2^2 \right).
\end{equation}
Note that because $h$ is a nonlinear function, the posterior
distribution is no longer Gaussian and so \eqref{posterior-mean-cov} no
longer applies. In this case, we turn to Markov Chain Monte Carlo (MCMC)
algorithms to generate samples from the posterior distribution and
compute the expectation of certain functions. A detailed discussion of
MCMC algorithms is outside the scope of this article and we refer the
reader to the monograph~\cite{casella} for an introduction to MCMC and
also to~\cite{somersalo} for applications of these algorithms to inverse
problems.

In this paper, we choose the preconditioned Crank-Nicolson (pCN)
algorithm of Cotter et al.~\cite{stuart-MCMC}, which is a variation of
the usual random walk Metropolis--Hastings algorithm.  Consider a
general setting where we would like to compute the expectation of some
function $f(\mb{v})$ under a Gaussian prior distribution of the form
$\pi_{\text{prior}}( \mb{v} ) = N(\mb{0},
\mb{C})$, and define
$$
\phi(\mb{v} , \mb{d} ) = \frac{1}{2} \left\| \pmb{\Sigma}^{-1}
  (\mb{F} (h(\mb{v}) - \mb{d}) \right\|_2^2.
$$
 Then we
have that
\begin{equation}
  \label{central-limit}
  \int f(\mb{v})
  \pi_{\text{posterior}}(\mb{v} \,|\, \mb{d}) \, d\mb{v} \approx
  \frac{1}{K} \sum_{k=1}^K \: f(\mb{v}^{(k)}), 
\end{equation}
where the samples $\mb{v}^{(k)}$ are generated by the following
algorithm:
\begin{enumerate}
\item Take $\mb{v}^{(0)} \sim \mb{N}(0, \mb{C})$.
\item While $k \le K$, 
  \begin{enumerate}[(i)]
  \item Generate $\mb{v} \sim N(\mb{0}, \mb{C})$.
  \item Set $\tilde{\mb{v}} = \sqrt{1 - \beta^2} \; \mb{v}^{(k)} +
    \beta \mb{v}$.
  \item With probability $a\left(\mb{{v}}^{(k)}, \mb{\tilde{v}} \right)
    = \min \left\{ 1, \, \exp \left[ \phi(\mb{v}^{(k)},\mb{d} ) -
 \phi(\tilde{\mb{v}}, \mb{d})\right]
    \right\}$, accept the sample and set 
    \begin{gather*}
      \mb{v}^{(k+1)} = \mb{\tilde{v}}.
    \end{gather*}
    Otherwise, reject the sample and set
    \begin{gather*}
      \mb{v}^{(k+1)} = \mb{v}^{(k)}.
    \end{gather*}    
\item $k \rightarrow k+1$.
  \end{enumerate}
\end{enumerate}
The parameter $K$ must be chosen large enough to ensure that the Markov
chain converges and that the error in the approximation of the integral
\eqref{central-limit} is sufficiently small. The parameter $0<\beta<1$
is also user-specified and controls the rate of convergence of the
algorithm.  In general, smaller values of $\beta$ result in a larger
acceptance probability in step 2(iii) of the above algorithm. This means
that the algorithm accepts more samples but the resulting Markov chain
will make correspondingly smaller jumps; therefore, when $\beta$ is
small, the iteration will be slower to traverse the posterior
distribution. If $\beta$ is large, then the Markov chain employs longer
jumps which is more desirable for exploring the parameter space;
however, the acceptance probability is also reduced which means that we
will reject most of the samples and occasionally get stuck at certain
points.  In practice it is desirable to choose $\beta$ so that the
acceptance probability from step 2(iii) lies in the range $[0.25,
0.35]$, which is close to the optimal acceptance rate of the random walk
algorithm~\cite{casella}. Finally, we note that the above algorithm can
be used to sample using non-centered priors by applying a simple linear
shift of $\mb{v}$ in the definition of $\phi$.

Once the samples are generated, we can store them in memory and use
them to estimate the posterior mean and standard deviation via
\begin{equation}
  \label{posterior-mean-postivie} 
  \mb{{v}}_{\text{PM}} := \int \mb{{v}} \: \pi_{\text{posterior}} (
  \mb{v} \,|\, \mb{d} ) \, d \mb{v} \qquad \text{and} \qquad
  \mb{C}_{\mb{v}} := \int (\mb{v} - \mb{v}_{\text{PM}}) (\mb{v} -
  \mb{v}_{\text{PM}})^T \: \pi_{\text{posterior}} ( 
  \mb{v} \,|\, \mb{d} ) \, d\mb{v}.
\end{equation}
We then define
\begin{equation}
  \label{posterior-mean-emission-positive}
  \mb{{q}}_{\text{sp}} := h( \mb{{v}}_{\text{PM}})
  \qquad\text{and}\qquad
  \mb{C}_{\text{sp}} := \int (h(\mb{v}) - \mb{q}_{\text{sp}}) (h(\mb{v})
  - \mb{q}_{\text{sp}})^T\: \pi_{\text{posterior}} (
  \mb{v} \,|\, \mb{d} ) \, d\mb{v},
\end{equation}
and take $\mb{{q}}_{\text{sp}}$ as our estimate of the non-negative
emission rates, where the subscript ``sp'' stands for ``smoothness and
positivity''.

To summarize, \eqref{MLE-estimate} defines the posterior mean
$\mb{q}_{\text{c}}$ under the prior assumption of constant emissions
which is used in \eqref{posterior-mean-cov} to derive the posterior mean
$\mb{q}_{\text{s}}$ under the smoothness assumption. This estimate is
then used in \eqref{posterior-mean-postivie} and
\eqref{posterior-mean-emission-positive} to derive the posterior mean
$\mb{q}_{\text{sp}}$ when both smoothness and positivity assumptions are
imposed.

% \todo{[What is $d$ in the above two equations?  What does CM stand for?
%   I'm not so keen on $()^\ast$ for transpose when you can use $()^T$
%   instead.  Or you've defined the Kronecker product, so why not just use
%   that instead \dots\ that's maybe best since I've screwed this all up
%   now and left you stuck with row vectors :-)  I'm sorry if you end up
%   having to reintroduce the transposes again. ]} 

\section{An industrial case study} 
\label{sec:case-study}

We now apply the methodology developed in the previous two sections to
an industrial case study in which airborne particles, most notably lead,
are emitted from a lead-zinc smelter operated by Teck Resources in
Trail, British Columbia, Canada.  These particulate emissions are of a
type called ``fugitive emissions'' that are not easily identified
because they are caused by accidental releases or come from buildings or
other areas of the industrial operation that are not amenable to direct
measurement approaches.  It is therefore of particular interest to
identify the location and amount of emitted material from such fugitive
sources in order to prioritize capital intensive projects to control
sources.

\begin{figure}[tbhp]
  \centering
  \includegraphics[width=0.8\textwidth]{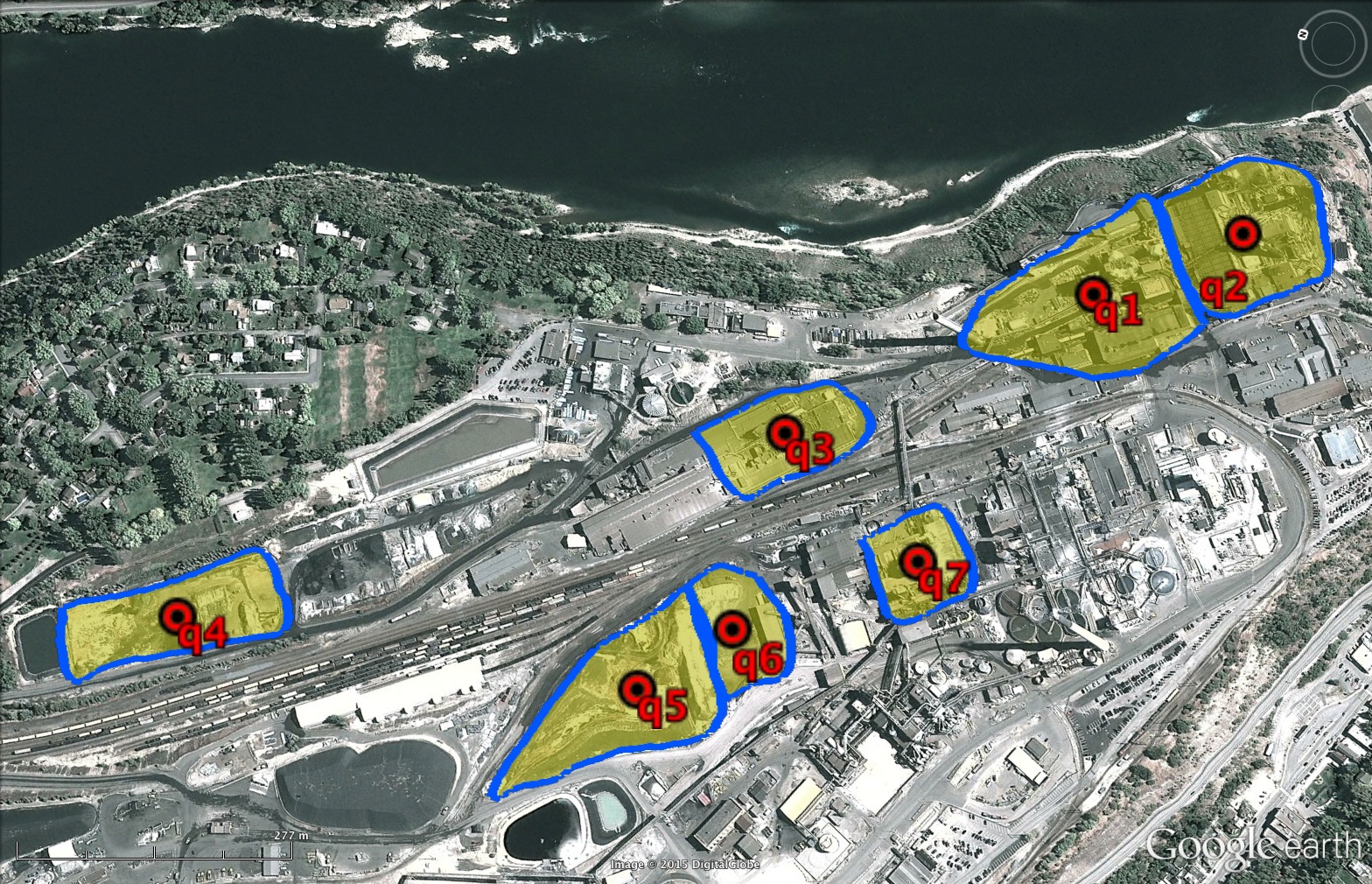}
  \caption{The smelter site in Trail, BC, highlighting the mostly likely
    areas to be sources of fugitive emissions.  The centroid of each
    area source is indicated by a point and labelled q1--q7.} 
  \label{fig:sources} 
\end{figure}

This study was performed over the period August 20 to September 19,
2013.  The company had installed sensors on the main stacks and other
non-fugitive sources that were already known to be historically major
emitters of lead particulates. Prior to the date of this study, fugitive
dust emissions reported by the company to Environment Canada's National
Pollutant Release Inventory (NPRI) were based on engineering
calculations listed in the NPRI reporting Toolbox (EPA method 42, which
does not account for local conditions). Figure~\ref{fig:sources} shows
an overhead view of the Trail smelter site along with certain areas that
were selected by the company's environmental engineering team as the
most likely sources of fugitive emissions (and the centroid of each area
is also identified).

\begin{figure}[tbhp]
  \centering
  \begin{tabular}{ll}
    (a) & (b)\\
    \includegraphics[height=0.2\textheight]{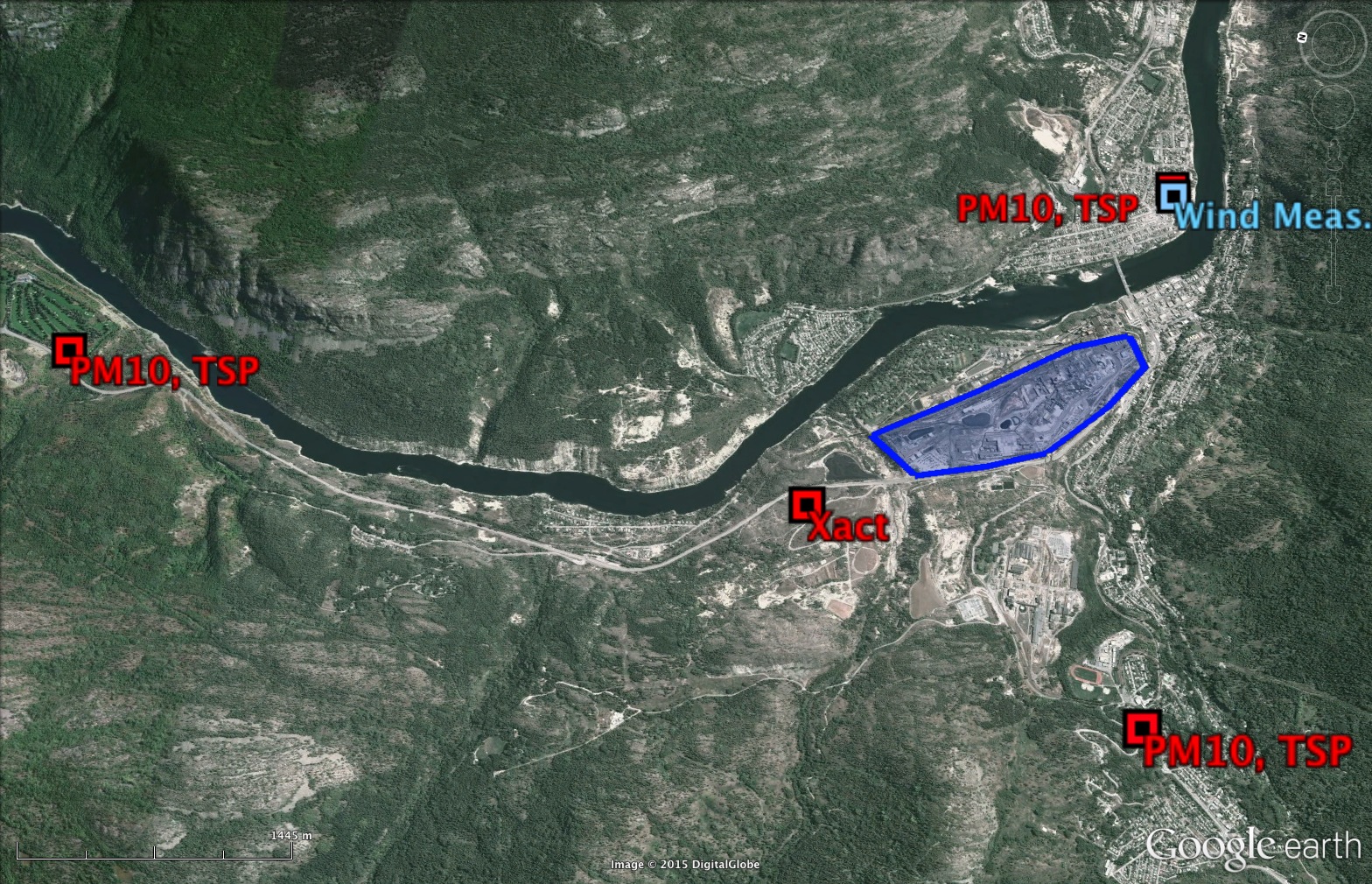}    
    & \includegraphics[height=0.2\textheight]{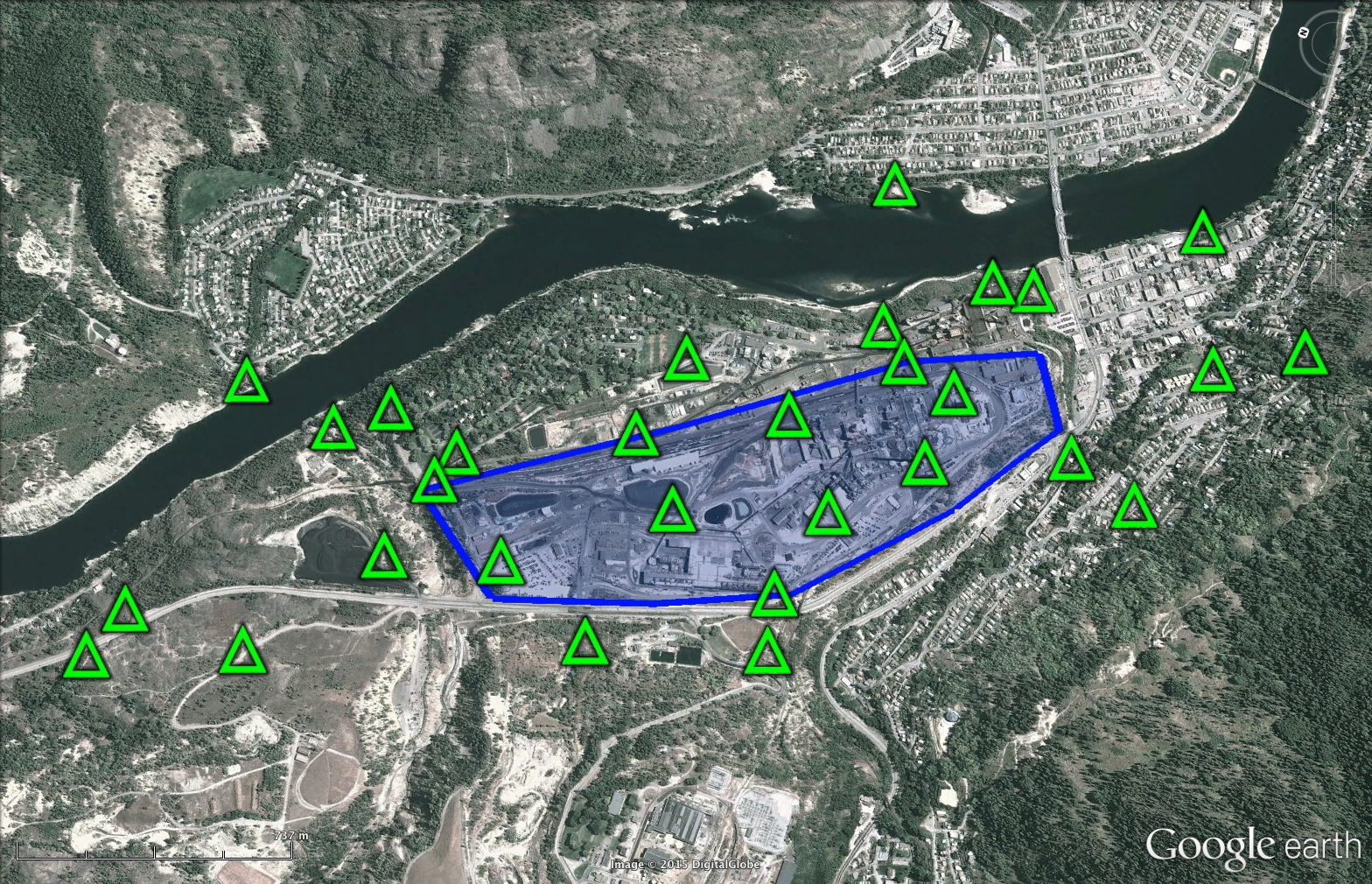}    
  \end{tabular}
  \caption{(a) Locations of the real-time measurement samples (Xact,
    PM10 and TSP systems) and the meteorological post where wind
    measurements were obtained, relative to the smelter site (outlined
    in blue). The TSP and PM10 measurements are taken with the same
    Andersen hi-vol device.  (b) A zoomed-in view of the outlined
    smelter site depicts the dust-fall jar locations with green
    triangles.}
  \label{fig:sensors}
\end{figure}

We were provided with meteorological data consisting of measurements of
wind velocity and direction at 10-minute intervals throughout the entire
one-month period.  There were also several types of data related to
particulate material obtained from a number of measurement devices, each
with a different sampling schedule:
\begin{itemize}
\item ``Dust-fall'': a set of 30 dust-fall jars deployed throughout the
  site, which yield measurements of the total deposited mass of lead
  (and various other particulates) accumulated over the entire monthly
  sampling period.
\item ``Xact'': real-time PM10 measurements (particulate material up to
  10\;$\myunits{\mu m}$ in diameter), which were taken using an Xact~620
  ambient metals monitor~\cite{cooper-environmental-2015} and reported 
  as hourly-averaged concentrations;
\item ``TSP'': total suspended particulate measurements (up to
  100\;$\myunits{\mu m}$ in diameter), taken with an Andersen high-volume
  air sampler~\cite{thermo-scientific-2015} as hourly-averaged
  measurements once every second day at midnight;
\item ``PM10'': the subset of the Andersen hi-vol TSP measurements that
  correspond to PM10 particulates (up to 10\;$\myunits{\mu m}$), taken as
  hourly-averaged measurements but only recorded once per week.
\end{itemize}
The locations of the dust-fall jars and real-time sensors are indicated
in Figure~\ref{fig:sensors}, and Table~\ref{tab:sensor-info} summarizes
the type of measurements each sensor provides. The sensor accuracy is
characterized using signal to noise ratio (SNR), which is the ratio of
the variance of the signal to that of the noise. The given values in
Table~\ref{tab:sensor-info} are chosen based on discussions with the
environmental engineering team at Teck Trail Operations.

\begin{table}[tbhp]
  \caption{Measurement characteristics for each
    sensor type.  %\todo{[Can you clarify what you mean by
    % ``data points per device''?  Is this what's available during the
    % one-month sampling period?]}
  } 
  \label{tab:sensor-info}  
  \centering
  \begin{tabular}{c|cccc}\hline
%    Symbol   & $\Delta$ &  s1, s3, s5 & s2, s4, s6 & s7   \\ 
    Type     & Dust-fall & TSP & PM10 & Xact  \\ 
    Schedule & monthly & 12:00am, every 2 days & 12:00am, weekly & hourly \\ 
    SNR & 10 &  100 & 100
    & 100 \\
    \# measurements/mo.& 1 & 16 & 6 & 761  \\ \hline 
  \end{tabular}
\end{table}

\subsection{Parameters and wind data}
\label{sec:params}

In this section, we summarize the input data specific to the Teck case
study that pertains to wind measurements and physical parameters for
lead particulates.  For simplicity, we restrict our attention to a
single particulate type corresponding to the most abundant form of lead
found by the sensors: lead monoxide (PbO).  A chemical analysis
performed by the company suggested that PbO particulate material has an
average diameter of $5\times 10^{-6} \:\myunits{m}$
\cite{EnvDustfallParticle} which we use as the size of all particulates
in this study. This value lies within the range of
0.8\,--\,20$\times10^{-6} \:\myunits{m}$ that is reported in
\cite{Owen-particlesize} for lead dust.  Based on this value of particle
size, we may then estimate the deposition velocity $W_{\text{dep}}$
using the data provided in~\cite[Figure~2]{slinn-deposition-velocity}.
Our assumed values of physical parameters for PbO are summarized in
Table~\ref{tab:physical-parameters}.
% and we will use these parameters in the forward model via equations
% \eqref{robin-bc}, \eqref{stokeslaw},\eqref{ermak-solution} and finally
% \eqref{forward-linear-form}.

\begin{table}[tbhp]
  \centering
  \caption{Physical parameters for lead monoxide (PbO), the dominant
    lead particulate encountered in this study.}   
  \label{tab:physical-parameters}
  %\scriptsize
  \begin{tabular}{c|cccc}
    \hline
    Property & Density & Diameter & Deposition velocity &
    Settling velocity \\
    Symbol & $\rho$          & $d_p$            & $W_{\text{dep}}$ & $W_{\text{set}}$ \\
    Units  & ${kg\: m^{-3}}$ & ${m}$            & ${m\: s^{-1}}$   & ${m\: s^{-1}}$\\ 
    Value  & $9530$          & $5\times 10^{-6}$& 0.005            &  0.0026 \\  \hline
  \end{tabular}
\end{table}

\begin{figure}[tbhp]
  \centering
  \begin{tabular}{c@{}c}
    \includegraphics[height=0.2\textheight]{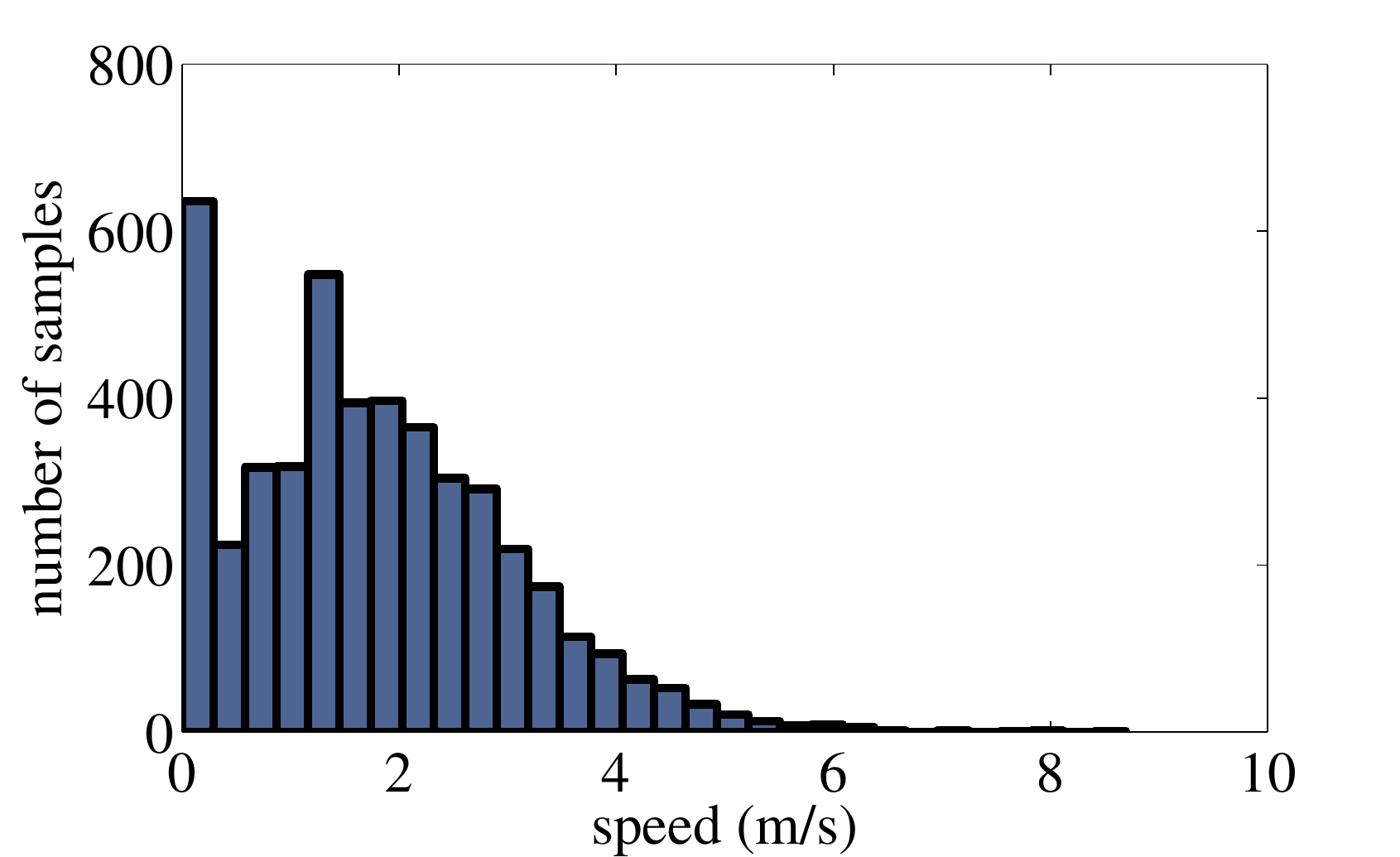}
    &
    \includegraphics[height=0.2\textheight]{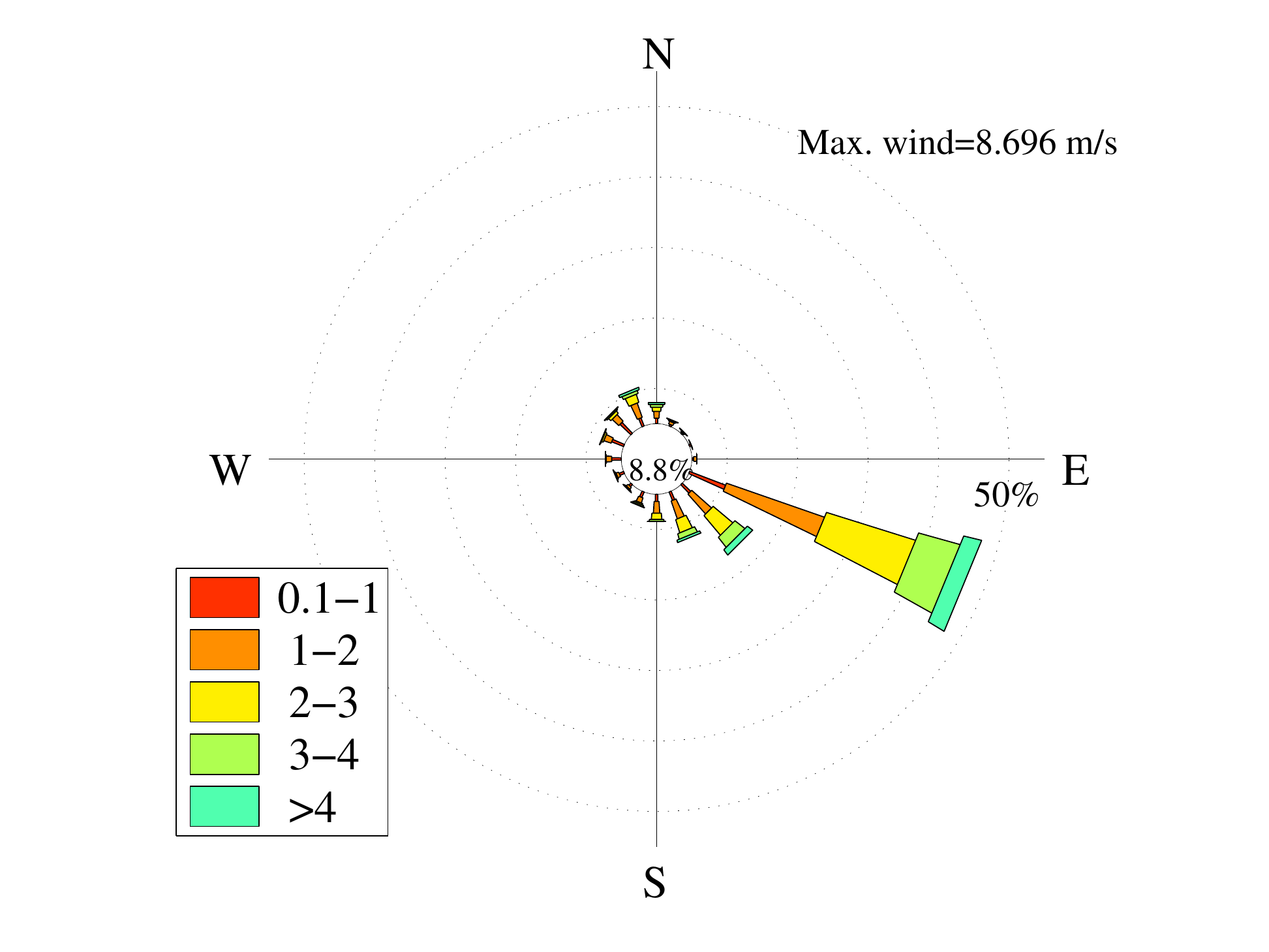}    
  \end{tabular}
  \caption{Wind speed histogram (left) and wind-rose diagram (right) for
    the period August 20--September 19, 2013.  The compass direction in
    the wind rose diagram denotes the direction that the wind is blowing
    \emph{from}.} 
  \label{fig:wind-data}
\end{figure}

\begin{figure}[tbhp]
  \centering
  \begin{tabular}{c@{}c}
    \includegraphics[height=0.15\textheight]{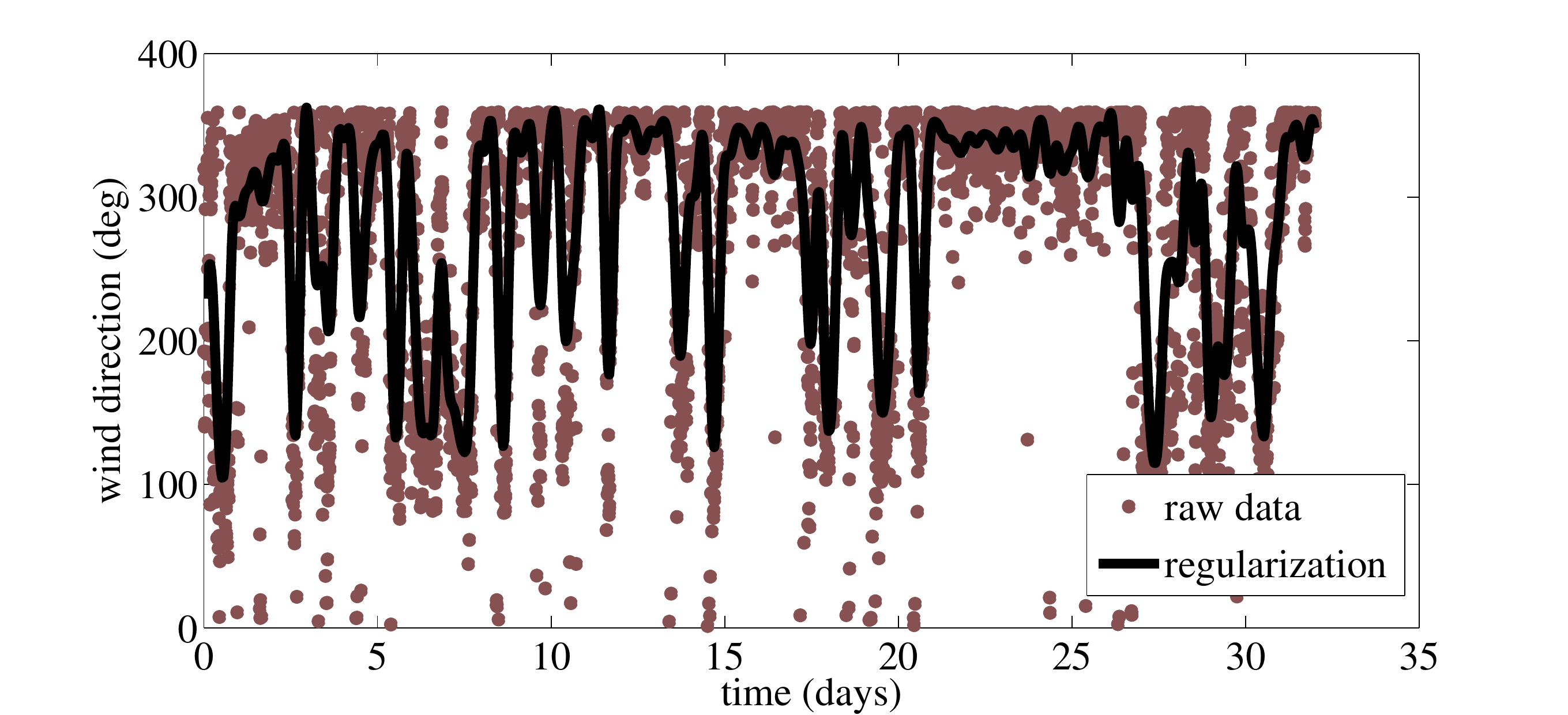}
    & 
    \includegraphics[height=0.15\textheight]{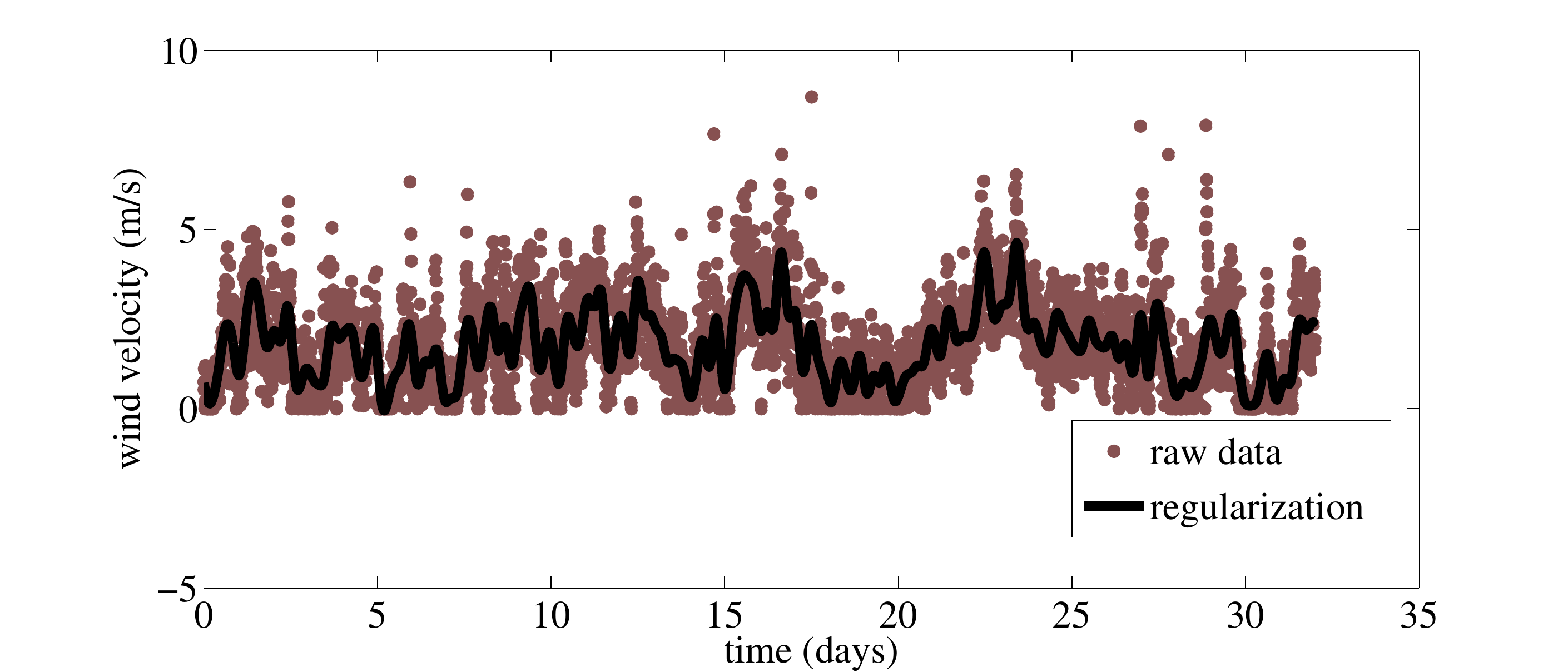}    
  \end{tabular}
  \caption{Comparison of regularized wind direction and velocity (solid lines)
    and raw measurements (points).}
  \label{fig:wind-regularization}
\end{figure}

The other essential inputs to the model are the wind velocity and
direction as functions of time. As mentioned earlier, wind data is
available at 10-minute intervals from a single meteorological station
identified in Figure~\ref{fig:sensors}a, while
%\todo{[Can you
 % mark this on the map in Fig.~2a?  I'm fairly certain it's near the
 %% PM10/TSP at upper right.  I remember it being near a park with a ball
 % diamond, and on Google Maps I think I can identify it as being at/near
 % 1815 2 Ave., coordinates $(49.096288,
 % -117.697890)$.]}.
Figure~\ref{fig:wind-data} presents the speed histogram and wind-rose
diagram corresponding to the month of interest.  The wind-rose data
indicate that the wind during this period blows consistently toward the
south-east, although there are significant variations in wind speed.
The raw wind data contains substantial measurement errors and as a
result they cannot be input directly into our forward model.  Instead,
we must first apply some form of regularization.  For this purpose, we
take the $y$-axis to point towards north and split the wind data into
velocity components in the $x$- and $y$-directions.  We then treat the
wind coordinates as independent variables and fit a Gaussian process
separately to each coordinate data set.  The fit is done by taking a
Gaussian covariance operator for each process with the variance of the
operator determined using ten-fold cross-validation.  The mean of the
Gaussian process obtained for each data set is then used as the
regularized wind input data for the forward model, and the resulting
wind components are pictured in Figure~\ref{fig:wind-regularization}.
Since this regularization procedure is based on a standard approach in
machine learning, we refer the interested reader to~\cite[Ch.~2, Ch.~3
and Sec.~5.3]{rasmussen} and~\cite[Sec.~6.4]{bishop} for details on the
use of Gaussian processes in regression.

\subsection{A simple test with synthetic data}
\label{sec:synthetic-data}

Before presenting results with the measured data, we start by testing
our framework on a synthetic data set in order to validate the method
and to also gain some insight into what type of reconstructions we can
expect from our algorithm.  The main challenge to be dealt
with here, and particularly in a real industrial setting, is that the
data is sparse and so one cannot expect to reconstruct emission rates
with high fidelity.

%\todo{[I'm unclear about what wind data you use for this synthetic test
%  case]}

\subsubsection{Generating the test data}

We consider the artificial emission rates pictured in
Figure~\ref{fig:sol-synthetic}a,b, which are imposed at seven sources
numbered q1--q7.  Each emission rate is a simple sinusoidal function of
time with a different amplitude and frequency, and with a positivity
constraint imposed if necessary.  Our motivation in making this choice
is to have time-varying emission rates that incorporate both high and
low frequencies, and which also have significant variations in
magnitude.  To generate the synthetic deposition measurements, we apply
the actual regularized wind data shown in
Figure~\ref{fig:wind-regularization} and discretize the forward problem
with uniform time increments of size $\Delta t =
1800\;\myunits{s}$. Note that this step size enters all aspects of the
forward problem including the Gaussian plume solution, wind
regularization, and measurement operators.  After solving the forward
problem \eqref{robin-bc}, \eqref{stokeslaw}, \eqref{ermak-solution} and
\eqref{forward-linear-form} to generate the artificial deposition
measurements, we introduce noise artificially by perturbing the data by
a vector of independent Gaussian random numbers having standard
deviations equal to the measurement errors listed in
Table~\ref{tab:sensor-info}.
 
\subsubsection{Solving the inverse problem}

\begin{figure}[tbhp]
  \centering
  \begin{tabular}{r@{}rr@{}r}
    \raisebox{0.17\textheight}{a)}
    & \includegraphics[height=0.2\textheight]{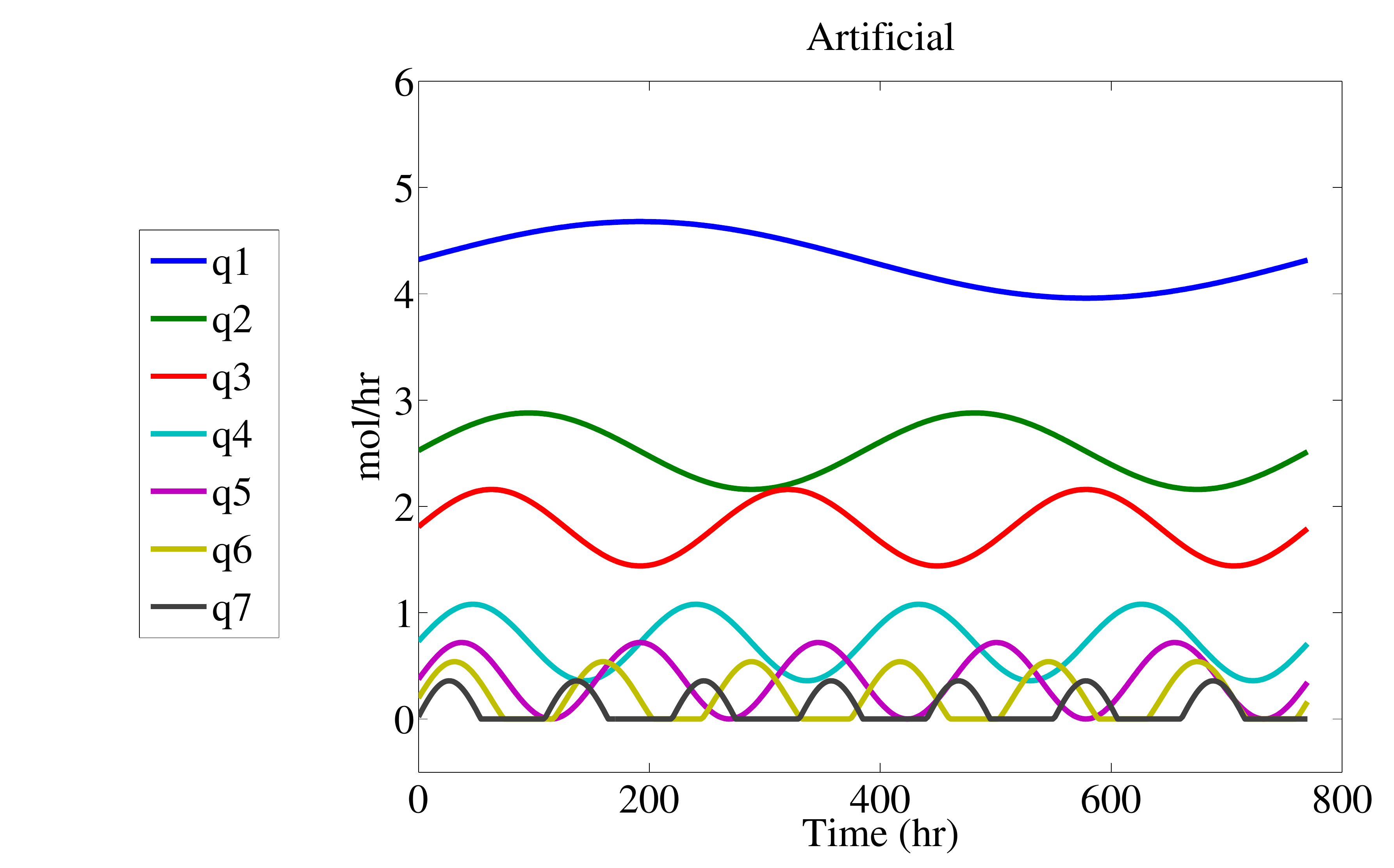}
    & \raisebox{0.17\textheight}{b)} & 
    % \begin{subfigure}{0.40 \textwidth}
    %   \includegraphics[height=4cm, clip=true, trim=3cm 0cm 0cm 0cm]{./figs/constant}
    % \end{subfigure}
    \includegraphics[height=0.2\textheight, clip=true, trim=3cm 0cm 0cm 0cm]{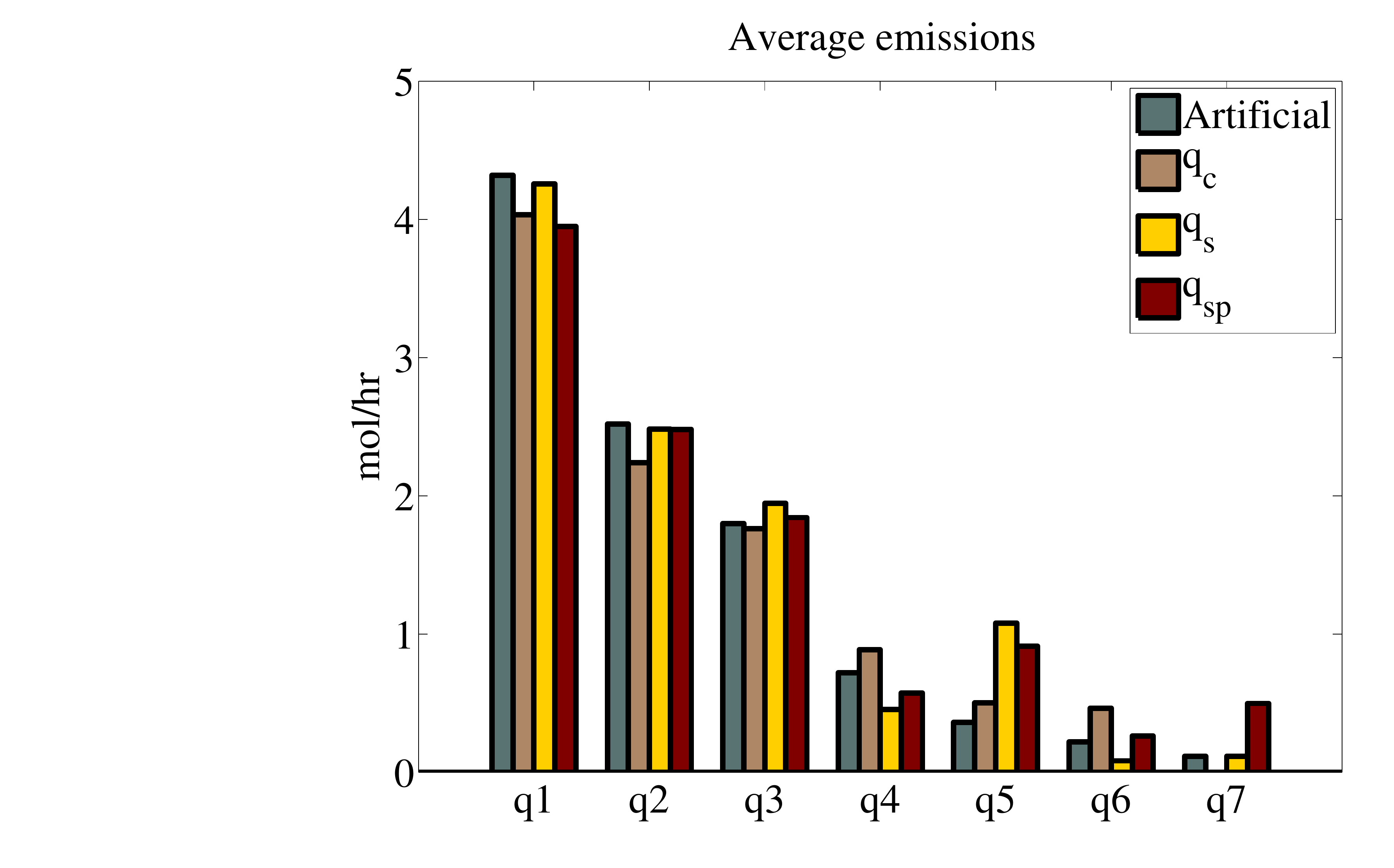}
    \\
    \raisebox{0.17\textheight}{c)} 
    & \includegraphics[height=0.2\textheight, clip=true, trim=3cm 0cm 0cm 0cm]{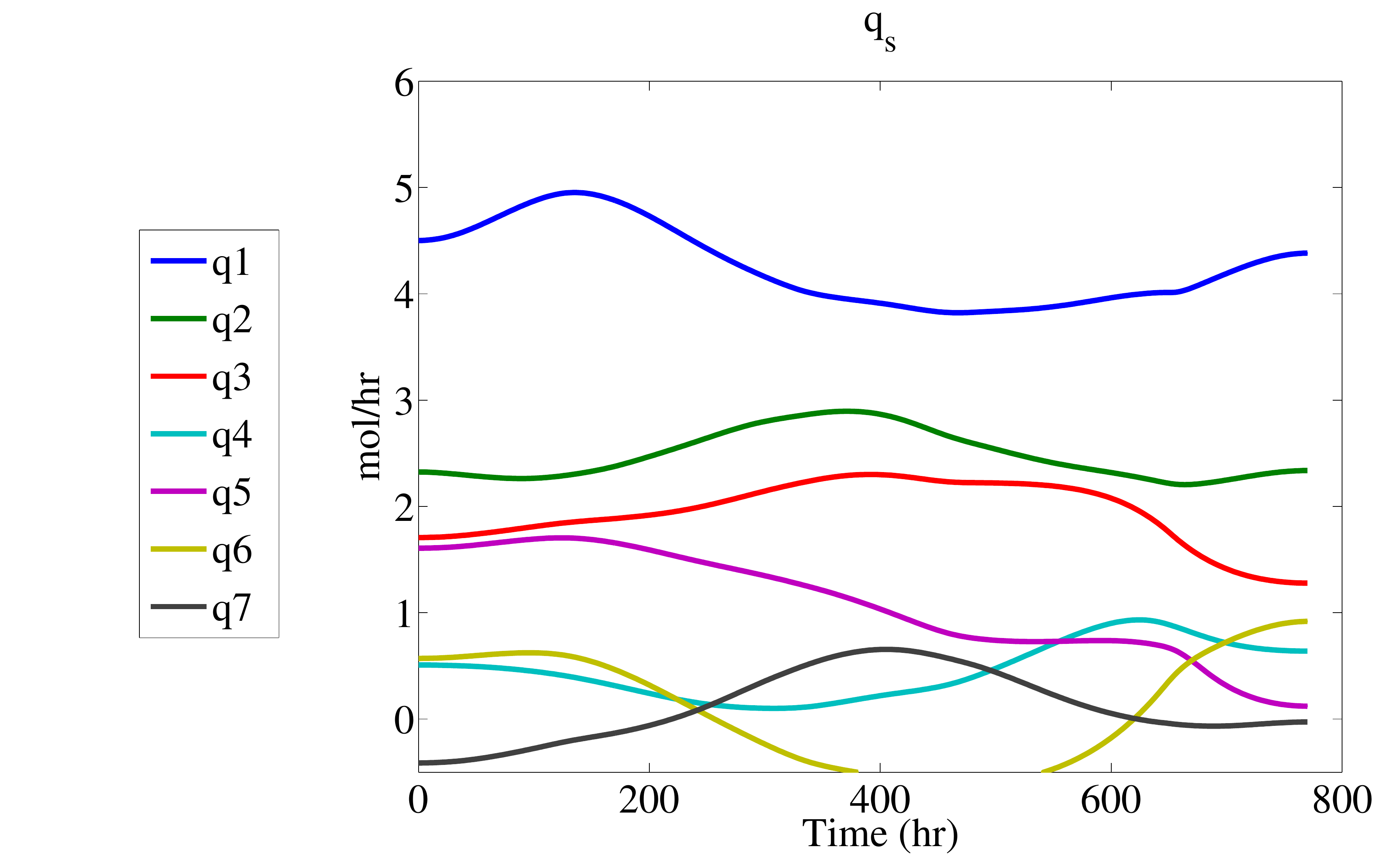}
    & \raisebox{0.17\textheight}{d)} 
    & \includegraphics[height=0.2\textheight, clip=true, trim=3cm 0cm 0cm 0cm]{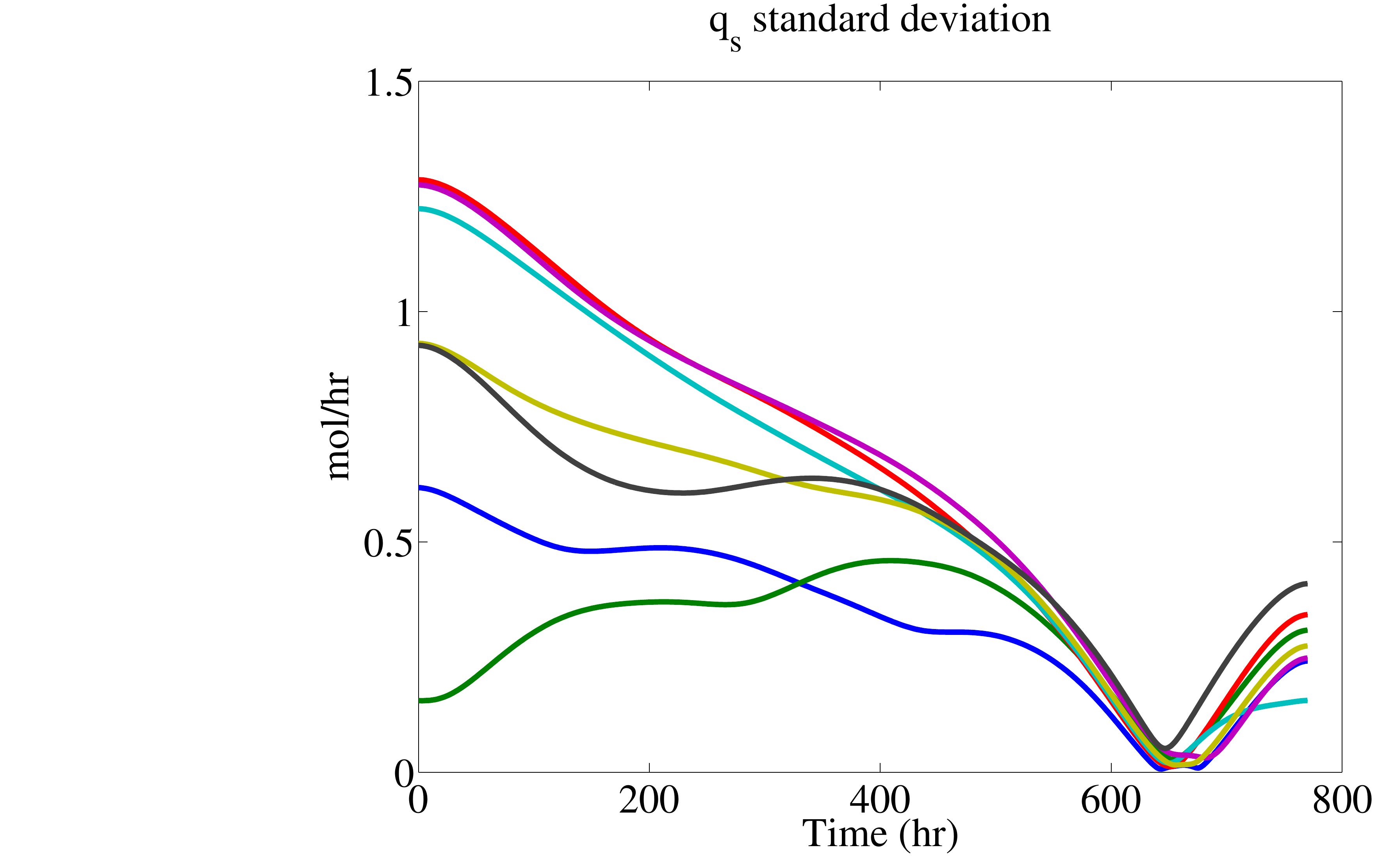}
    \\
    \raisebox{0.17\textheight}{e)} & 
    \includegraphics[height=0.2\textheight, clip=true, trim=3cm 0cm 0cm 0cm]{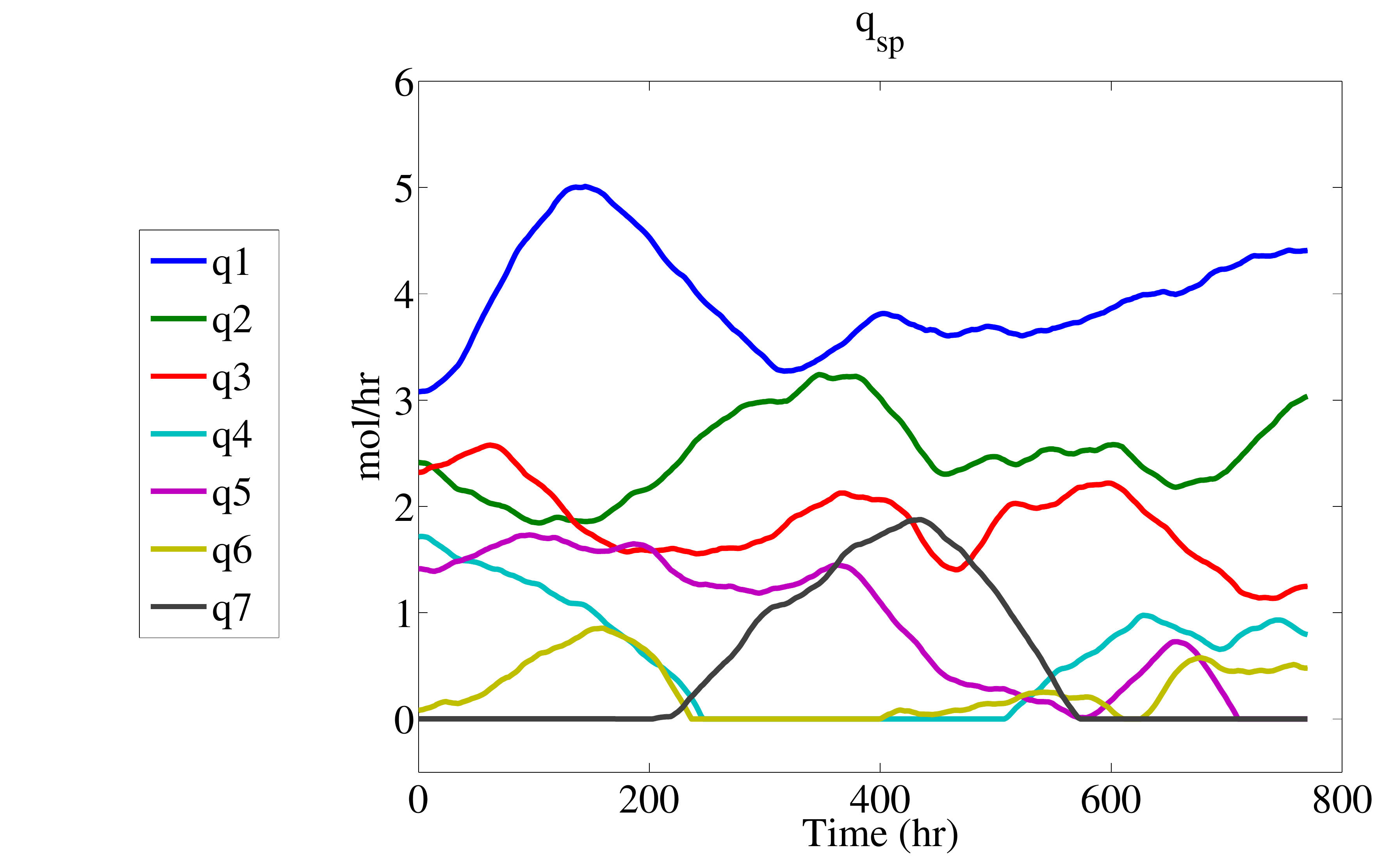}
    & \raisebox{0.17\textheight}{f)} 
    & \includegraphics[height=0.2\textheight, clip=true, trim=3cm 0cm 0cm 0cm]{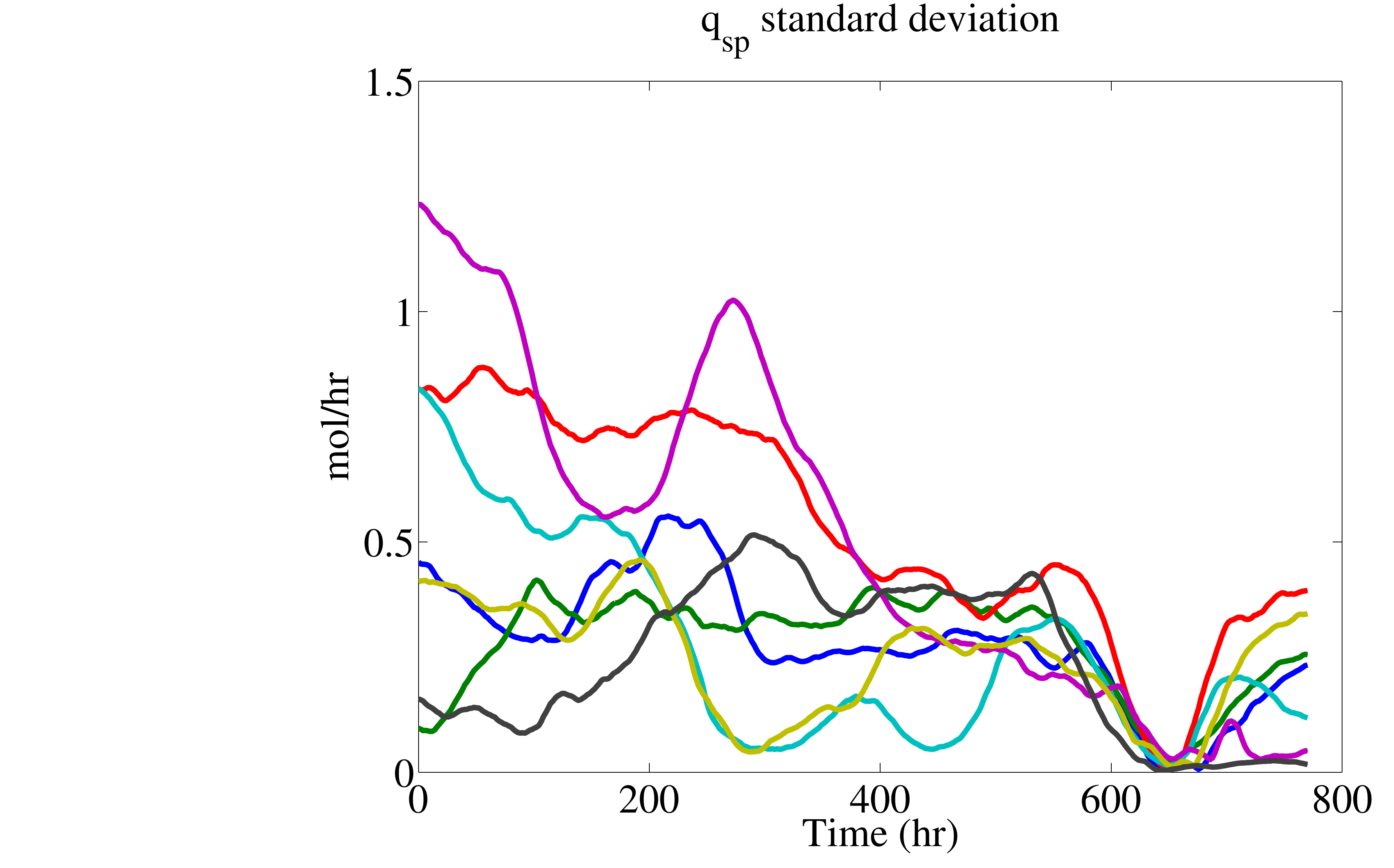}
  \end{tabular}
  % \\
  % e)
  % f)
  % \begin{subfigure}{0.40 \textwidth}
  %   \includegraphics[height=4cm, clip=true, trim=3cm 0cm 0cm 0cm]{./figs/total}
  % \end{subfigure}
  \caption{Summary of results using the artificial input data set. (a)
    Artificial emission rates used to generate the data. (b) Average
    emission rate of each source obtained using the three different
    priors for constant emissions, smooth emissions and smooth and
    positive emissions. (c) Posterior mean of sources as functions of
    time using the smoothness assumption only (no positivity constraint)
    along with (d) the standard deviations. (e) Posterior mean of
    estimated emissions with the both smoothness and positivity
    assumptions along with (f) corresponding standard deviations.}
  \label{fig:sol-synthetic}
\end{figure}
In order to avoid the ``inverse crime'' of choosing the same mesh to
generate the artificial data and solve the inverse
problem~\cite{somersalo}, we instead solve the inverse problem using a
coarser time increment $\Delta t = 3600\;\myunits{s}$, which ensures
that our wind data, forward map and measurement operators are different
from those used to generate the data.  We will also consider a setting
where the measurement noise is mis-specified and construct the
$\pmb{\Sigma}$ matrix using measurement errors equal to one-half the
value listed in Table~\ref{tab:sensor-info}.

We begin by solving the inverse problem under the assumption that
emission rates are constant, following the methodology outlined in
Section~\ref{sec:constant-emission}. The resulting solution
$\mb{q}_{\text{c}}$ is pictured in Figure~\ref{fig:sol-synthetic}b,
where the grey bars (labeled ``Artificial'') depict the average of the
emission rates in Figure~\ref{fig:sol-synthetic}a.  Our estimate
indicates that the constant emissions assumption is relevant if one is
interested in the time-averaged source emission rates, although no
information is provided about the time dependence.
% Of course this solution tells us
%nothing about variations of the sources in time. \todo{[I don't
 % understand this either.]}

%We now construct a prior distribution that can be used to solve the
%inverse problem using~\eqref{posterior-mean-cov} in 
We next turn our attention to the case of smooth (non-constant) emission
rates without a positivity constraint.
% \todo{[so you're not
 % applying positivity here]}.
We use the same noise covariance matrix $\pmb{\Sigma}$ as in the
constant case and substitute the vector $\mb{q}_c$ of
Figure~\ref{fig:sol-synthetic}b into \eqref{posterior-mean-cov} to
compute the posterior mean and covariance.  Results of this computation
are presented in Figures~\ref{fig:sol-synthetic}c,d, which depict the
posterior mean and standard deviation.  Note that the posterior mean is
not strictly a physically reasonable estimate of the emission rates
because one source (q6) exhibits negative emissions.  Nonetheless, our
algorithm still correctly identifies the largest sources, for which the
magnitude and overall shape of the emission curves for the larger
sources is captured quite well.  However, we note that the smaller
sources are not resolved by our reconstruction, which is apparent in the
standard deviation plots of Figure~\ref{fig:sol-synthetic}d.  By
considering the standard deviation as a measure of uncertainty in the
reconstructions, we see that the uncertainty is larger for the smaller
sources (q3--q7) and significantly smaller for the two main sources
(q1,\,q2), meaning that we should have less confidence in our
reconstruction for the smaller sources.  This is due to the scarcity of
the data as well as our choice of prior which prefers smooth estimates.

As a final test, we impose the positivity constraint using the approach
described in Section~\ref{sec:non-negative}. As mentioned before, there
is no longer an analytic expression for the posterior distribution and
so we employ an MCMC algorithm to sample the posterior and compute the
mean and standard deviations.  Figures~\ref{fig:sol-synthetic}e,f depict
the estimate of $\mb{q}_{\text{sp}}$ and its standard deviation using
$K=5\times 10^5$ and $\beta = 0.6$, which yields an average acceptance
probability of $0.32$. These estimates with the positivity constraint
appear slightly better than the ones without, which is reflected in the
smaller values of standard deviation.

% \subsubsection{Additional sensors}
%
% As was mentioned previously, the main difficulty in estimation of the
% emission rates in our setting is the scarcity of the data. This is
% apparent from the standard deviation plots of
% Figure~\ref{fig:sol-synthetic}. Which indicate that the variances are
% larger for $t \le 400$ and this is due to wind direction and position
% of the Xact sensor which is our most informative sensor. The company
% has shown interest in purchasing two new Xact sensors in the future
% and here we will address the question of positioning these sensors.

\subsection{Results for actual measured deposition data}
\label{sec:real-data}

\begin{figure}[tbhp]
  \centering
  \begin{tabular}{c@{}cc@{}c}
    \raisebox{0.17\textheight}{a)} & 
    \includegraphics[height=0.2\textheight]{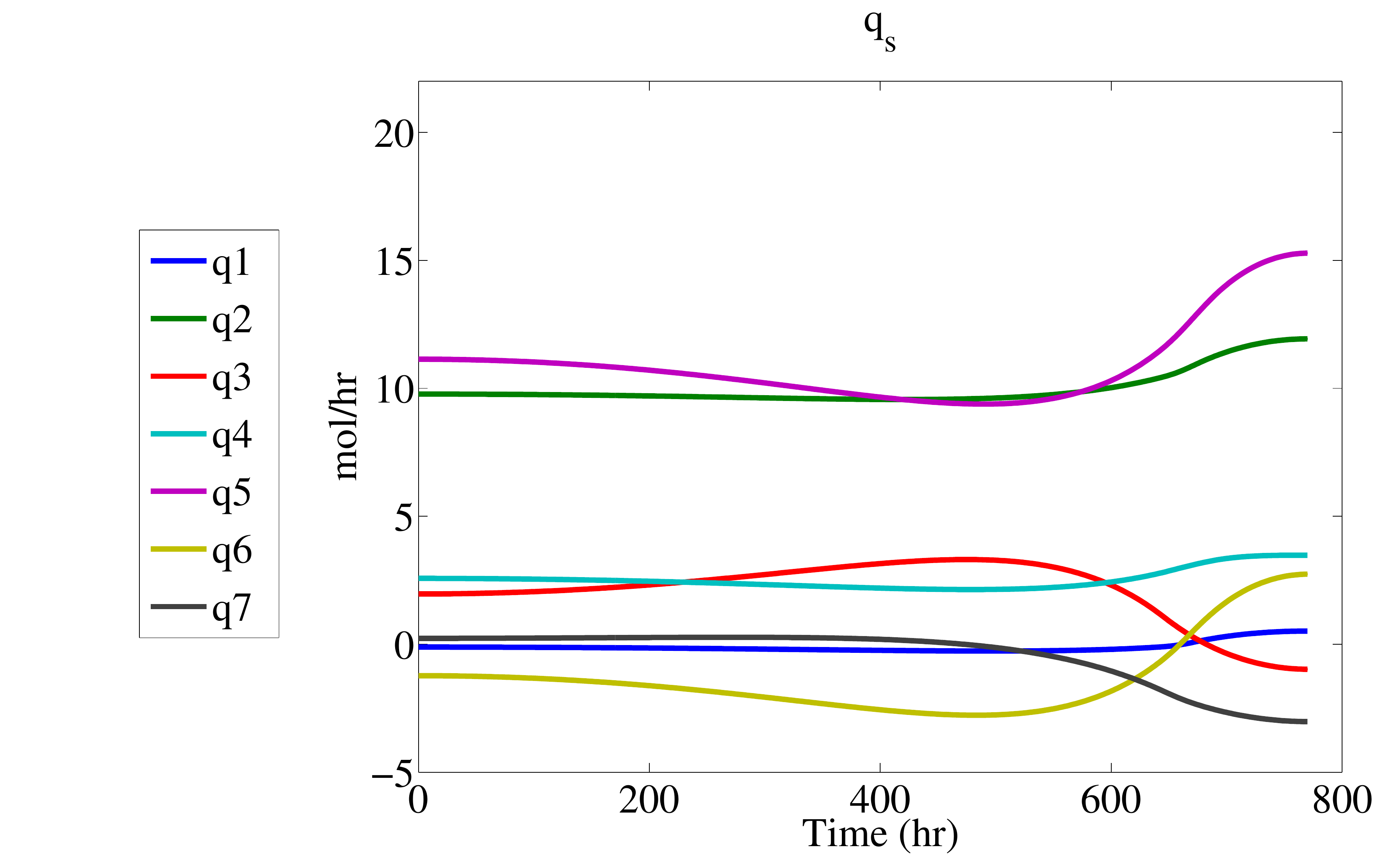}
    & \raisebox{0.17\textheight}{b)} & 
    % \begin{subfigure}{0.40 \textwidth}
    %   \includegraphics[height=4cm, clip=true, trim=3cm 0cm 0cm 0cm]{./figs/constant}
    % \end{subfigure}
    \includegraphics[height=0.2\textheight, clip=true, trim=3cm 0cm 0cm 0cm]{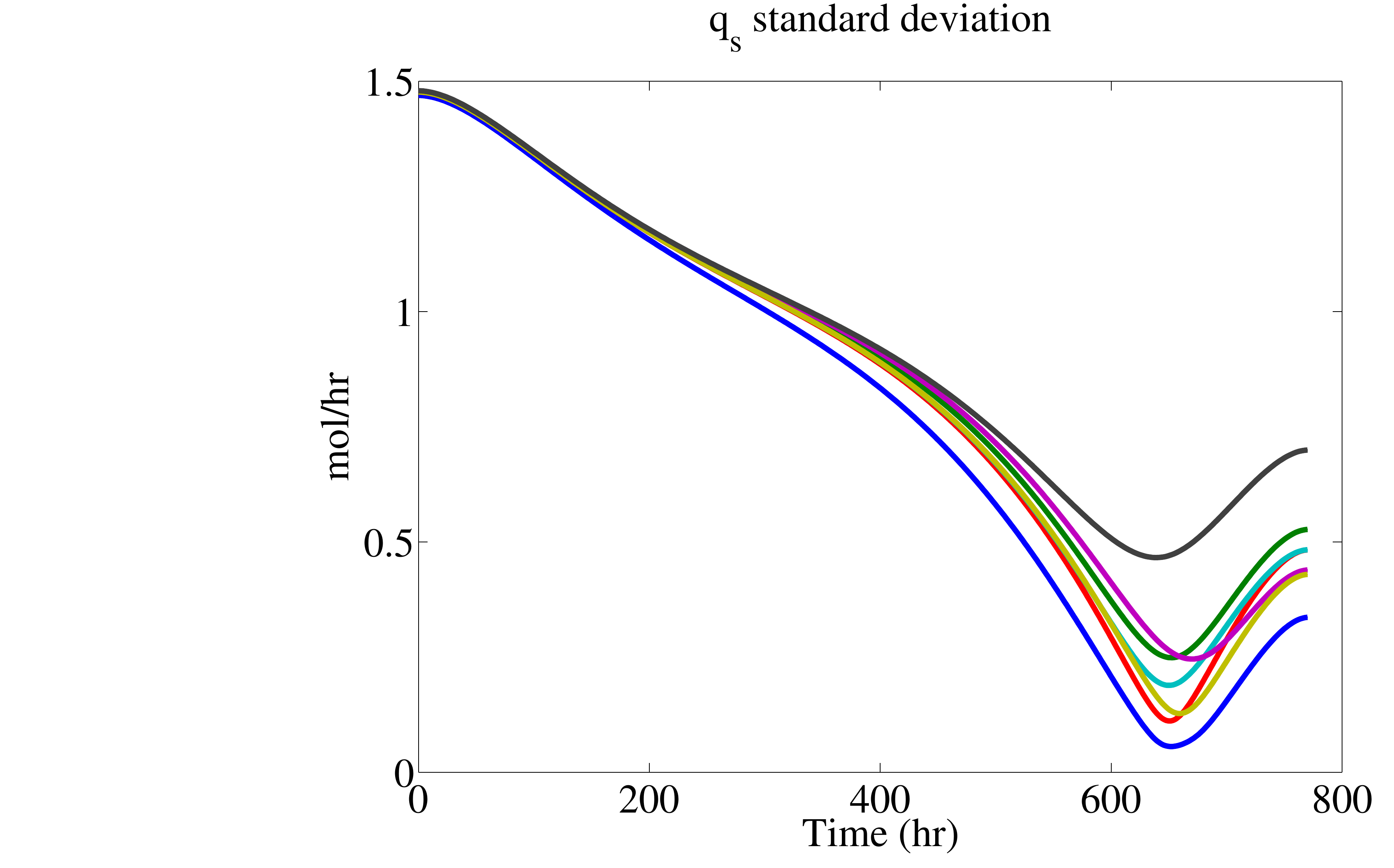}
    \\
    \raisebox{0.17\textheight}{c)} & 
    \includegraphics[height=0.2\textheight]{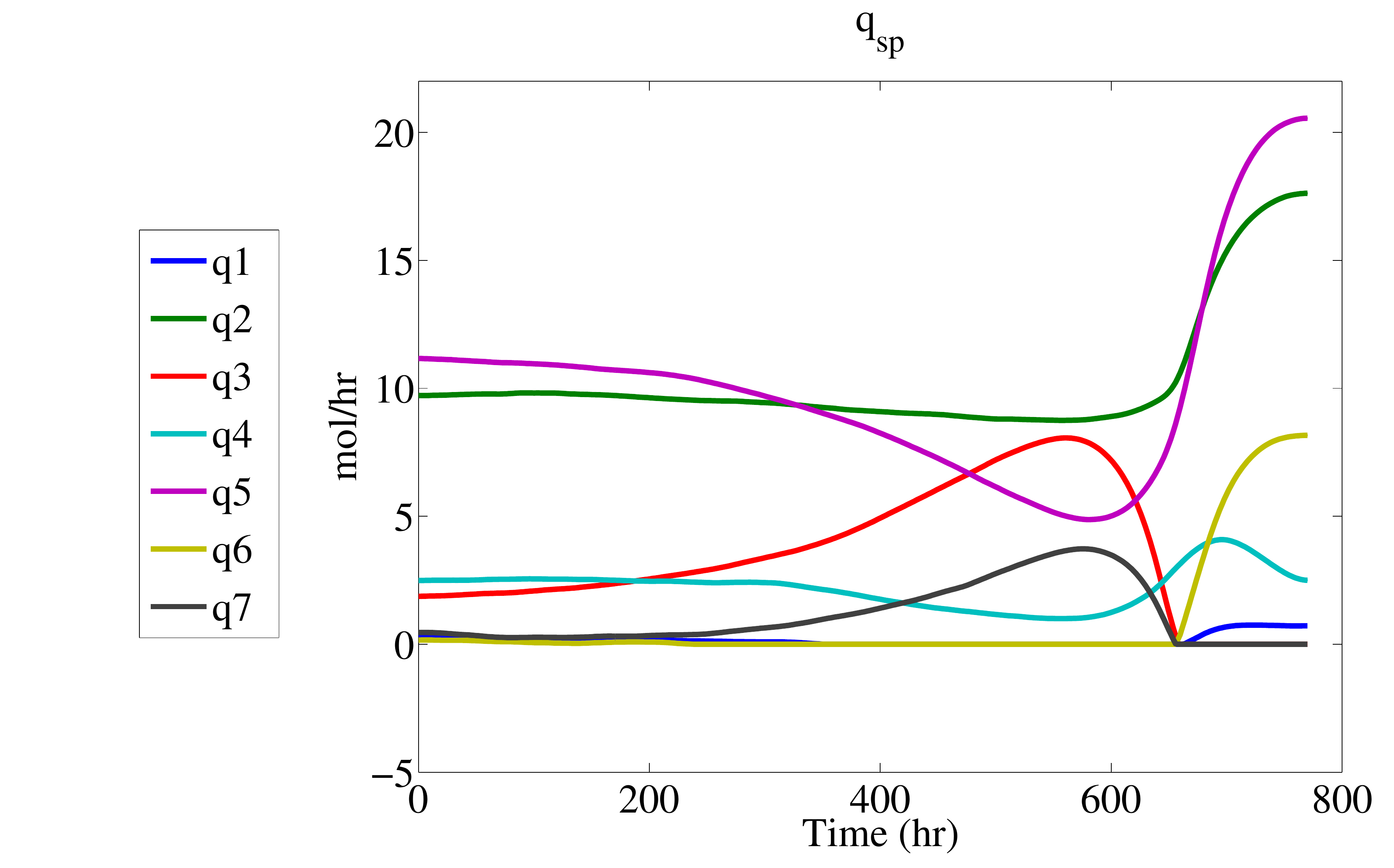}
    & \raisebox{0.17\textheight}{d)} & 
    \includegraphics[height=0.2\textheight, clip=true, trim=3cm 0cm 0cm 0cm]{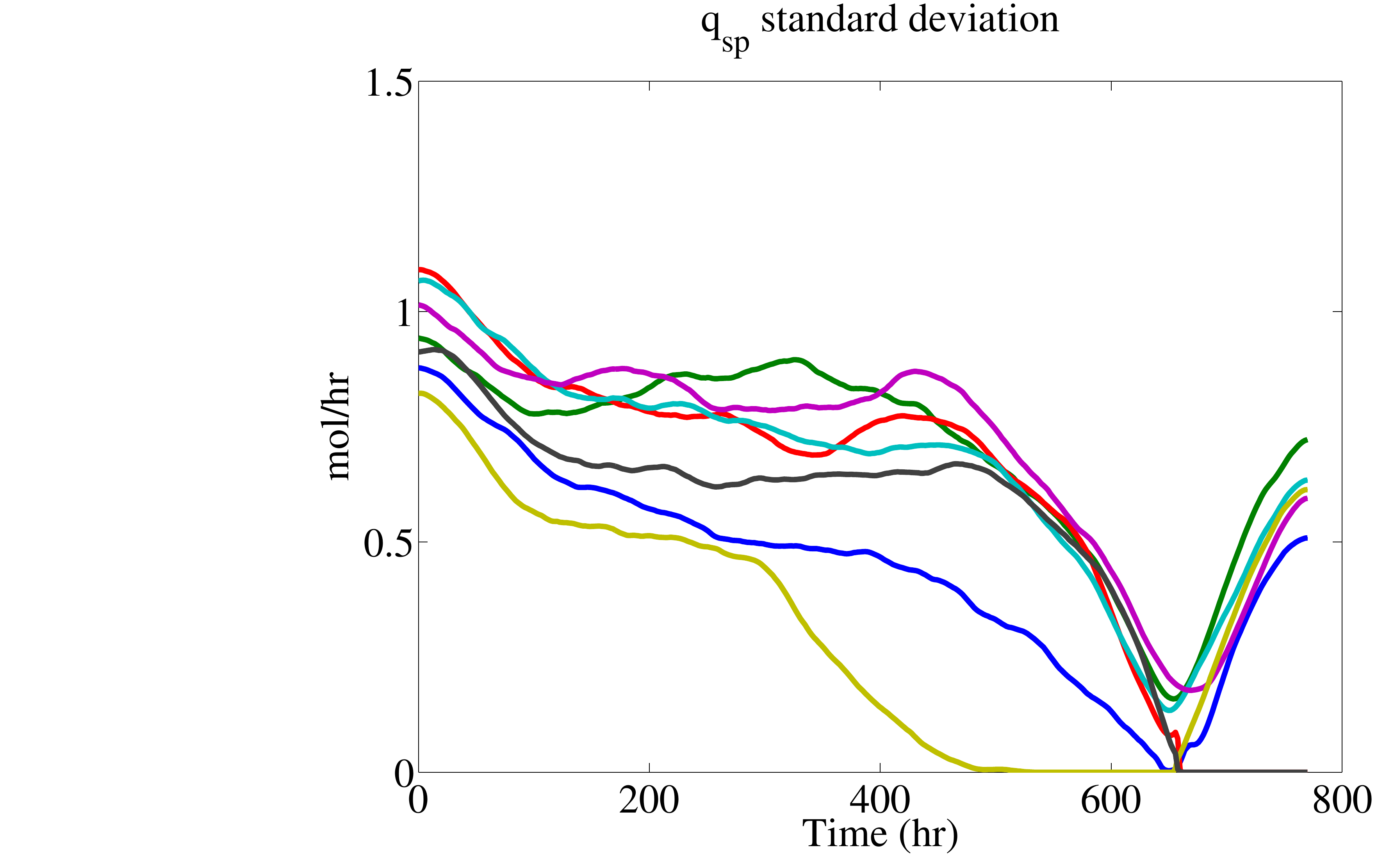}
    \\
    \raisebox{0.17\textheight}{e)} & 
    \includegraphics[height=0.2\textheight]{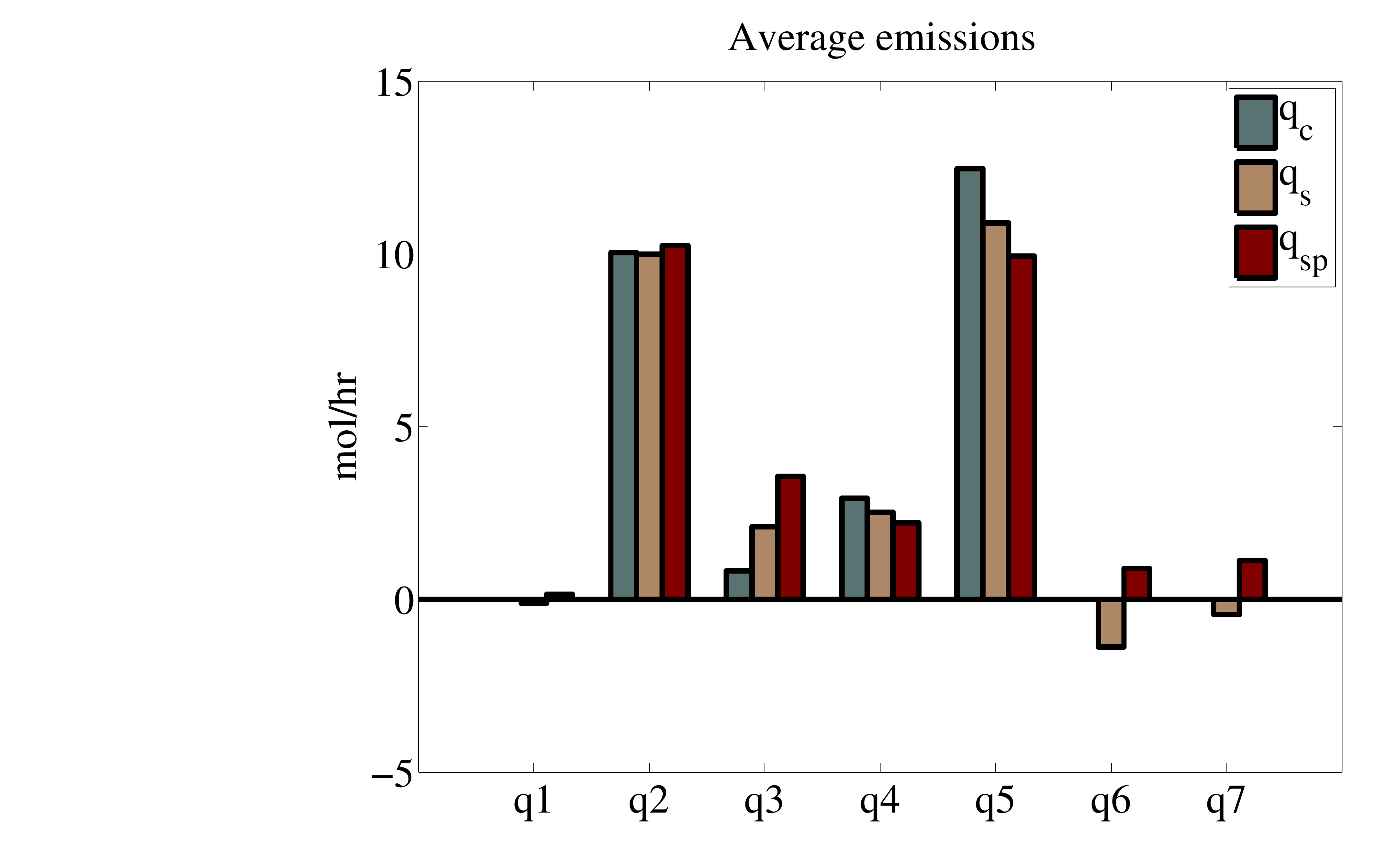}
    &\raisebox{0.17\textheight}{f)} & 
    \includegraphics[height=0.2\textheight, clip=true, trim=3cm 0cm 0cm 0cm]{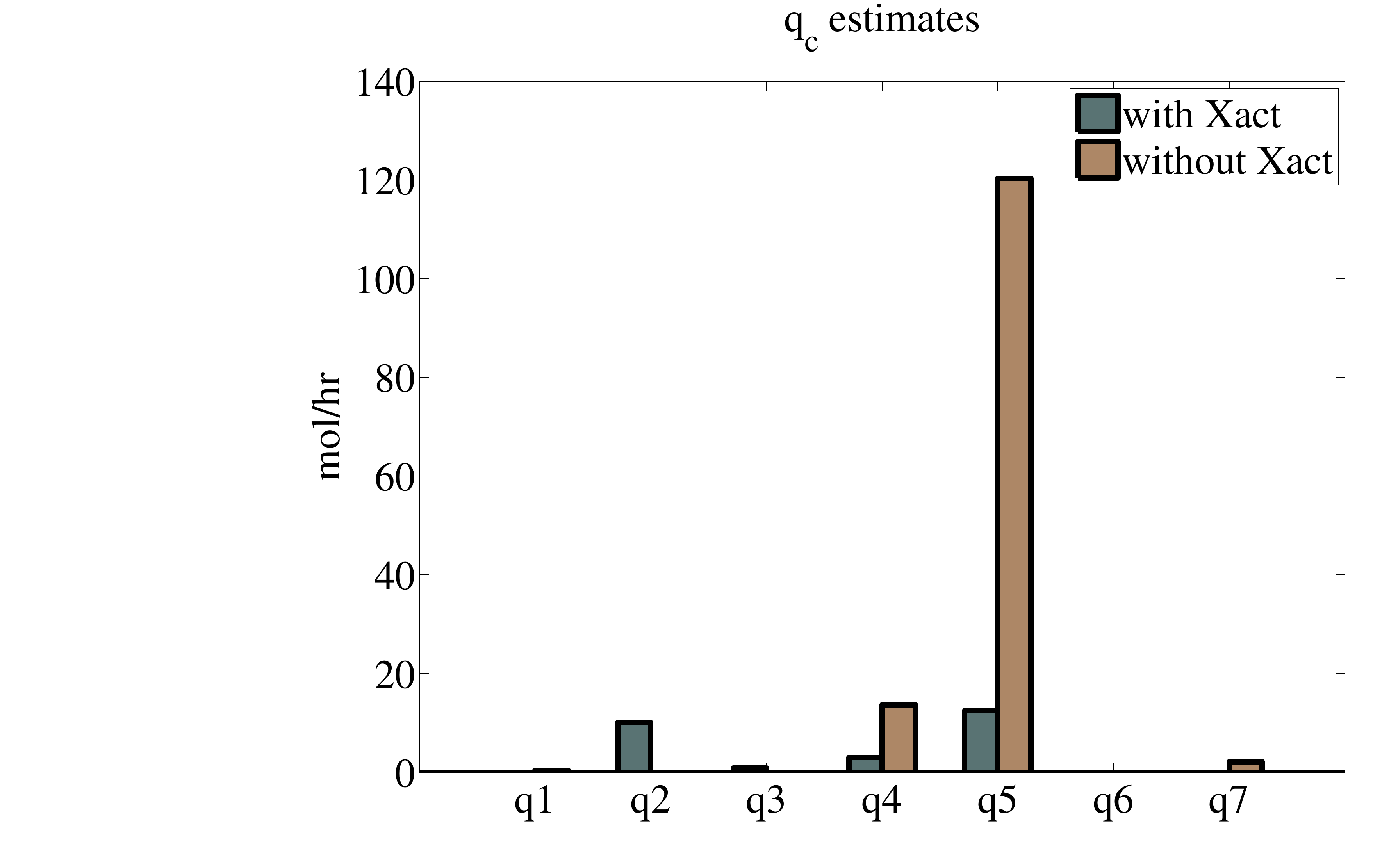}
  \end{tabular}
  \caption{Results obtained using actual measured depositions at the
    Trail smelter. (a,b) Posterior mean and standard deviation of the
    emissions using the smoothness prior with no positivity
    constraints. (c,d) Posterior mean and standard deviation obtained
    using the smoothness and positivity prior. (e) Average emission rate
    of each source for the duration of the study, using three choices of
    prior. (f) Comparison of the $q_c$ estimate obtained using the
    entire data set and the estimate when the Xact data is excluded.}
  \label{fig:sol-actual}
\end{figure}

We now turn our attention to solving the inverse problem using the
actual measured deposition values at the Trail smelter site.  We use the
same parameter values as for the artificial problem above and results
are summarized in Figure~\ref{fig:sol-actual}. The constant emissions
estimate in Figure~\ref{fig:sol-actual} suggests that q2 and q5 are by
far the main sources of lead particulates.  This is to be expected for
q5, since it is a loading area where piles of material are mixed and
stored and hence is expected to generate significant quantities of
airborne dust, whereas q2 is the area immediately to the north of the
smelter building.  On the other hand, q1, q6 and q7 are all predicted to
emit minimal amounts of lead. Figures~\ref{fig:sol-actual}a,c show the
posterior mean of the estimates using the smoothness prior with and
without a positivity constraint. Averaged emission rates pertaining to
these cases are presented in Figure~\ref{fig:sol-actual}e. The relative
importance of the various sources does not differ dramatically between
the three different algorithms, perhaps with the exception of the
estimates for q6 and q7.

It is interesting that Figures~\ref{fig:sol-actual}a,c both exhibit a
surge in certain emission rates (q2,\,q5,\,q6) near the end of the
month.  Looking at the standard deviation plots in
Figures~\ref{fig:sol-actual}b,d, we note that these surging month-end
values are estimated with a higher level of confidence, compared to the
beginning of the month when uncertainties are larger. Note also that the
uncertainties are smaller in Figure~\ref{fig:sol-actual}d as compared to
\ref{fig:sol-actual}b which indicates that imposing the positivity
constraint yields improved estimates.  In the next section, we will
discuss in more detail the uncertainties in solution estimates and
how they can be used to assess the impact of emissions on the
surrounding area.

Another insightful result of this case study is the sensitivity of the
solution to the different types of measurements. The Xact device is the
most expensive instrument, but it also provides nearly six times the
measurement data than all other instruments combined.  There are thus
clear advantages to having more Xact data, although deploying new Xact
devices may be difficult to justify due to their high cost and
maintenance requirements. Figure~\ref{fig:sol-actual}f depicts the
$\mb{q}_{\text{c}}$ estimates obtained with and without the Xact data,
clearly demonstrating that Xact measurements have a major impact on the
outcome. Without the Xact device, our approach greatly over-estimates
emission rate from source q5 and also fails to identify q2 as the
primary smelter building contributing to lead emissions.  This clearly
demonstrates the importance of having real-time measurements available
in such an emissions study.  Indeed, in the absence of time-varying data
from such a real-time sensor, one simply obtains a posterior that is
very similar to the prior and hence it is not possible to reconstruct
general time-varying sources with any degree of accuracy.
% As it is often the case with using actual data the inverse problem is
% more difficult to solve in this
% case. Figures~\ref{fig:sol-actual}b,d show the estimated posterior
% variance for each source, from which we observe that variances are
% quite large when $t \le 500$, especially in the case of $\mb{q}_s$
% \todo{[for ...]}. This indicates that the data is less informative
% during this initial time period and so the posterior variance has not
% been reduced significantly.  \todo{[However, after this intial
%   transient, blah,blah,blah, \dots\ what happens near $t=600$?  Can
%   you expand on the discussion here a little bit?]}

\subsection{Impact assessment and uncertainty propagation}
\label{sec:impact}

Now that we have obtained the solution to the inverse problem, it is
natural to go one step further and study the broader implications of the
emissions estimates in terms of how lead particulates are distributed
over the area surrounding the smelter site.  These effects can be
quantified using various measures such as total annual lead emissions or
monthly ground-level deposition.  The total annual emission of a given
pollutant is typically of great interest to a company such as Teck
because it forms part of their annual reporting requirements to
environmental monitoring bodies, not to mention that companies often set
ambitious goals for reducing the total emission figure.  It is
straightforward to extrapolate our average emission rates from
Figure~\ref{fig:sol-actual}e to estimates of the annual emission rate
for the entire smelter.  Table~\ref{tab:annual-emission} shows the
results of our computation along with annual emissions that were
given in an independent study that was performed by the company
\cite{thep-report}. The fact that our estimates are close to the
values in the independent study increases our confidence in the obtained solutions.
 The reported values of standard deviations were
obtained by first approximating the posterior with a Gaussian (this is
only needed in the case of $\mb{q}_{\text{sp}}$). We also present an
estimate of the $90\%$ probability interval, which is the interval
around the mean that contains $90\%$ of the probability mass. This
interval is also computed based on the Gaussian approximation to the
posterior.
\begin{table}[tbhp]
  \centering
  \caption{Extrapolated total annual emission rates (in
    $\myunits{tonne/yr}$) for lead particulates.  The standard
    deviation and $90\%$ probability intervals are computed by first
    estimating the posterior with a Gaussian.} 
  \label{tab:annual-emission}
  \begin{tabular}{cl l l}\hline 
    Approximation & $\mybunits{tonne/yr}$   & Standard deviation &
                                                                     90\%
                                                                     probability
                                                                     interval\\\hline
    $\mb{q}_{\text{c}}$  & $51.0$ & &\\
    $\mb{q}_{\text{s}}$  & $46.1$  & $5$ & $\pm 8.2$\\
    $\mb{q}_{\text{sp}}$ & $54.9 5$ &$5$&  $\pm 8.2$ \\
    Independent study \cite{thep-report} & $45$ & &  \\ \hline
    % Independent studies  & $15$ to $50$ \\\hline
  \end{tabular}
\end{table}

\begin{figure}[tbhp]
  \centering
  \begin{tabular}{cc}
    a)
    \includegraphics[width=0.45\textwidth, clip=true, trim=0.8cm 0cm 0.8cm 0cm]{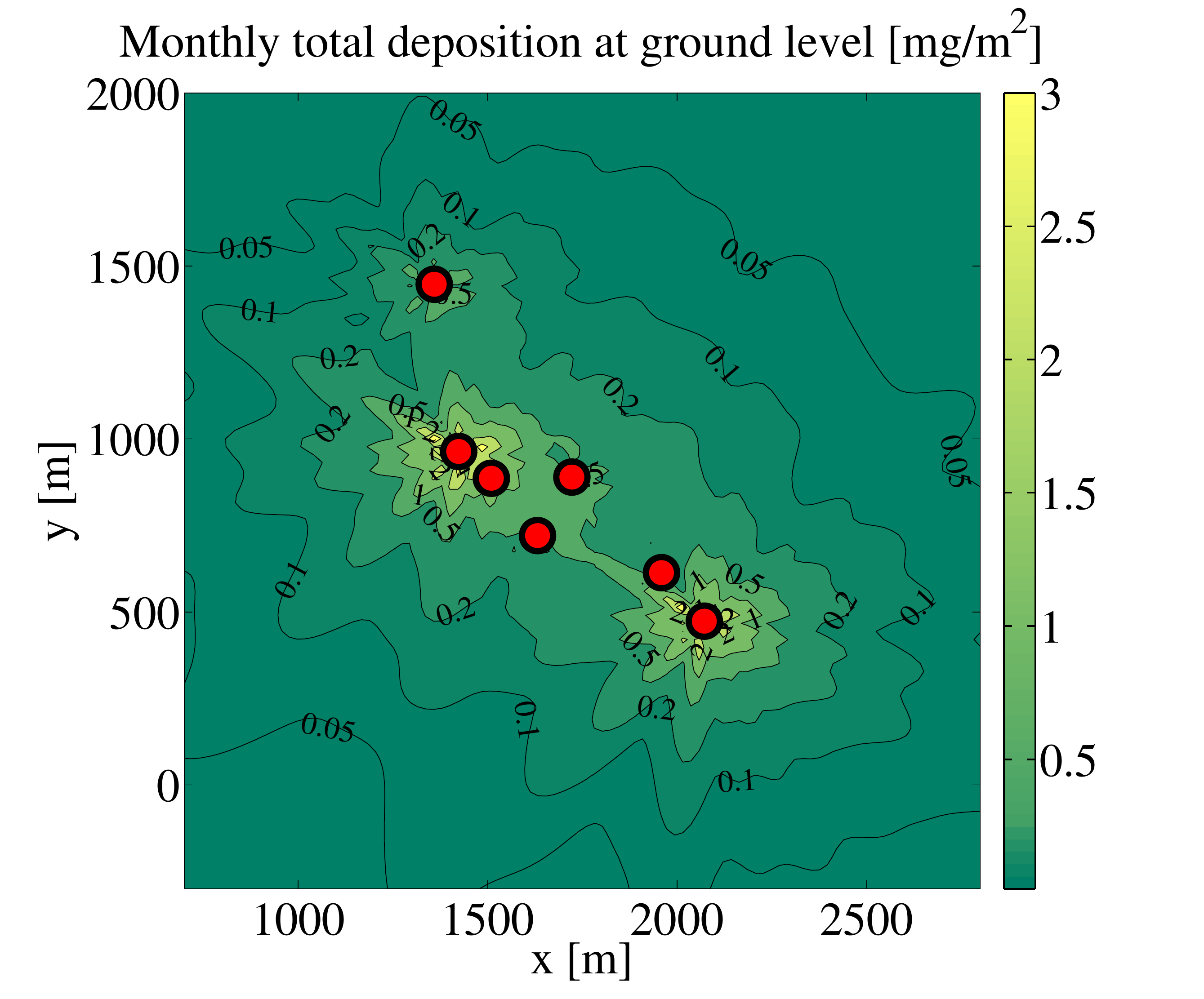} 
    &
    b) \includegraphics[width=0.45\textwidth, clip=true, trim=0.8cm 0cm 0.8cm 0cm]{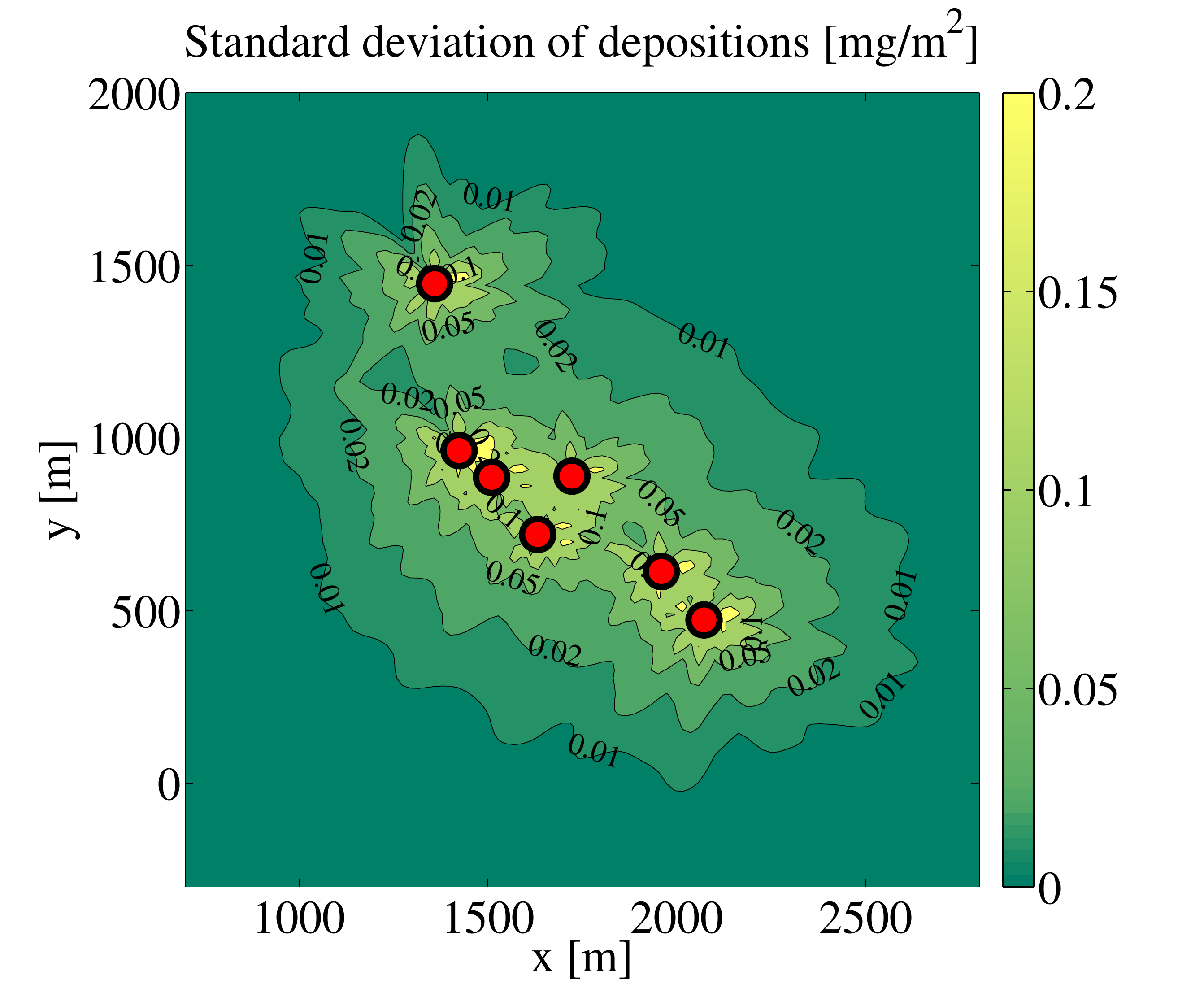}
  \end{tabular}
  \caption{(Left) Deposition in $\myunits{mg/m^2}$ obtained by solving
    the forward problem with the posterior mean
    $\mb{q}_{\text{sp}}$. (Right) Standard deviation of deposition using
    a low-rank approximation of the posterior covariance.  The solution
    is computed on a $100 \times 100$ spatial grid using only the first
    100 eigen-pairs.}
  % \todo{[The title on the std deviation plot is misleading, since it's
  %   very easy to miss the ``std'' at the end.  Can you replace the
  %   title with something like ``Standard deviation of deposition''?]}} 
  \label{fig:forward-uq}
\end{figure}

A more interesting problem is that of computing the total lead deposited
at ground level using our estimated emission rates for the period of
August 20 to September 19, 2013.  This is an important problem because
it allows us to assess in more detail the impact of emissions on the
area surrounding the industrial site. In order to perform this
computation we take a domain consisting of a rectangular ground
patch, $[L_x, U_x] \times [L_y, U_y] \subset \reals^2$, which is
discretized on an $n_x \times n_y$ grid of equally-spaced points.  We
use the Gaussian plume solution \eqref{ermak-solution} to compute lead
concentration at each grid point, which is then integrated in time and
multiplied by the deposition velocity to obtain monthly deposition
values. This procedure is similar to our construction of the $\mb{F}$
operator in \eqref{linear-measurement}.  We then let $\mb{b}$ denote the
vector containing grid point values of the monthly depositions and
define an operator $\mb{H}$ such that
\begin{equation} \label{forward-on-grid}
  \mb{b} = \mb{H} \mb{q} 
\end{equation}
for a given vector of emission rates $\mb{q}$. We can now solve the
forward problem using our estimate $\mb{q}_{\text{sp}}$ to obtain
$\mb{b}_{\text{sp}} = \mb{H} \mb{q}_{\text{sp}}$. The resulting solution
is depicted as a contour plot of lead mass per unit area in
Figure~\ref{fig:forward-uq}a, which indicates that most deposition
occurs close to the sources and within the boundaries of the
smelter site itself.
% \todo{[What does the contour plot of standard deviations show?]}

\begin{figure}[tbhp]
  \centering
  \includegraphics[height=0.2\textheight]{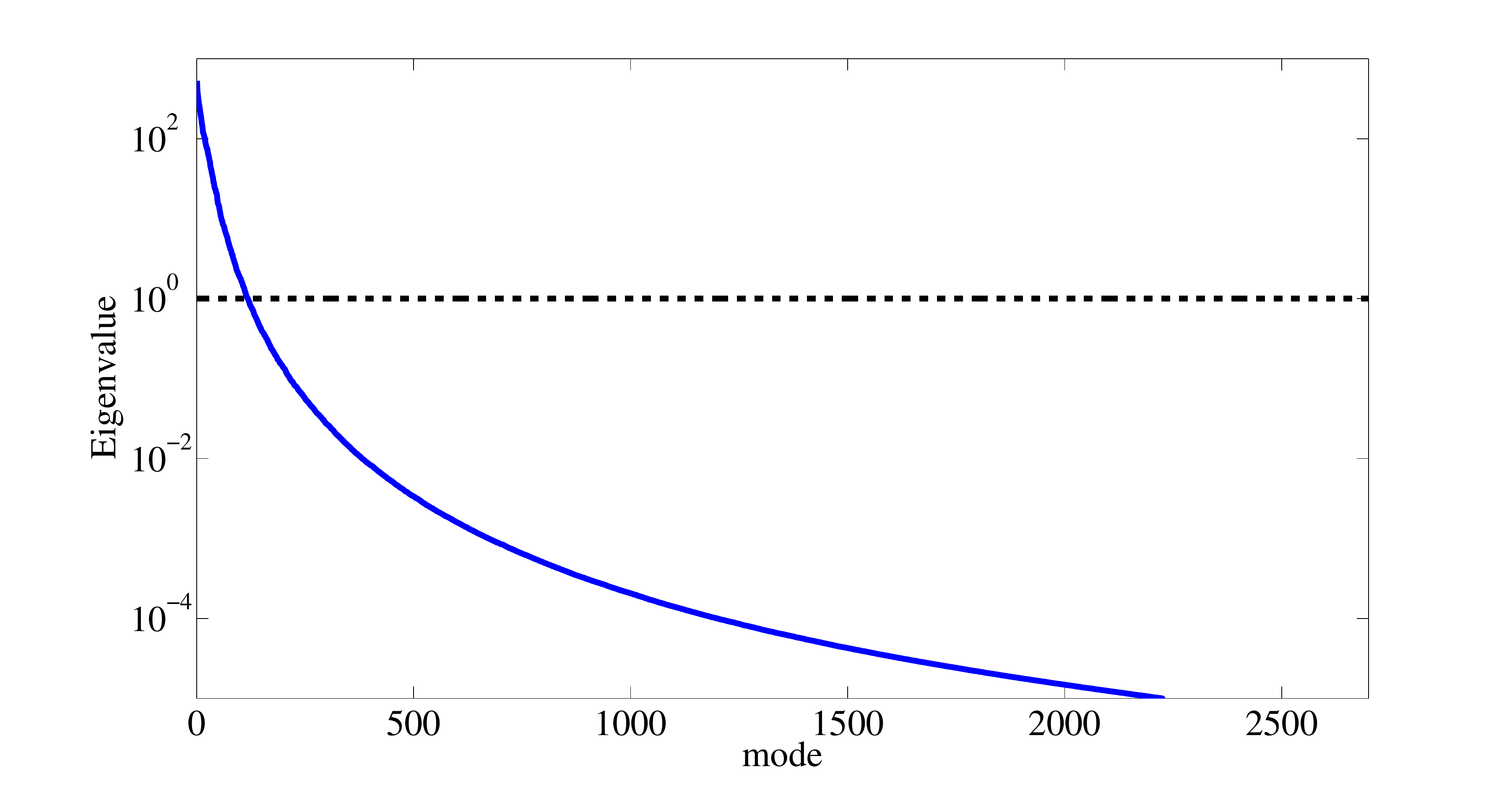}
  \caption{Eigenvalues of the posterior covariance for the solution of
    the inverse problem with positivity constraint.}
  \label{fig:eigenvalue}
\end{figure}

Next, we propagate the posterior uncertainty through the forward model
to obtain error bounds on the monthly deposition values, with our main
aim being to obtain a framework that provides a reasonable estimate of
solution uncertainty on a fine grid.  To make the computations as
efficient as possible, we would like to avoid sampling algorithms and so
we approximate the posterior distribution for the non-negative solution
using a Gaussian distribution $N(\mb{q}_{\text{sp}},
\mb{C}_{\text{sp}})$. Our aim is to exploit the fact that a linear
transformation of a Gaussian distribution is also Gaussian.  Recall that
$\mb{C}_{\text{sp}}$ can be estimated using the MCMC algorithm of
Section~\ref{sec:non-negative}.  Because~\eqref{forward-on-grid} is
linear, we have that $\mb{b}_{\text{sp}} \sim N( \mb{H}
\mb{q}_{\text{sp}}, \mb{H} \mb{C}_{\text{sp}} \mb{H}^T)$. Computing this
covariance matrix on a fine grid might still be intractable because
$\mb{H}$ is a large matrix and $\mb{C}_{\text{sp}}$ is dense.  However,
Figure~\ref{fig:eigenvalue} shows that the eigenvalues of
$\mb{C}_{\text{sp}}$ decay rapidly in time, which suggests replacing it
by a low-rank approximation by simply truncating the spectrum.  To this
end, let $\mb{C}_{\text{sp}} = \mb{L} \mb{D} \mb{L}^T$ be the usual
eigenvalue decomposition of the covariance, let $n_e$ be the number of
eigenvalues we want to retain, and take $\tilde{\mb{D}}$ as the first
$n_e\times n_e$ sub-matrix of $\mb{D}$. Then let $\tilde{L}$ denote the
tall matrix containing the first $n_e$ columns of $\mb{L}$.  We can now
define $\tilde{\mb{C}}_{\text{sp}} := \tilde{\mb{L}} \tilde{\mb{D}}
\tilde{\mb{L}}^T$ and approximate the distribution of the depositions as
$\mb{b}_{\text{sp}} \sim N( \mb{H}\mb{q}_{\text{sp}}, \mb{H}
\tilde{\mb{C}}_{\text{sp}} \mb{H}^T)$, and the covariance can now be
estimated using $n_e$ solves of the forward problem.

\begin{figure}[tbhp]
  \centering
  \begin{tabular}{r@{}rr@{}r}
    \raisebox{0.17\textheight}{a)}  
    & \includegraphics[height=0.2\textheight]{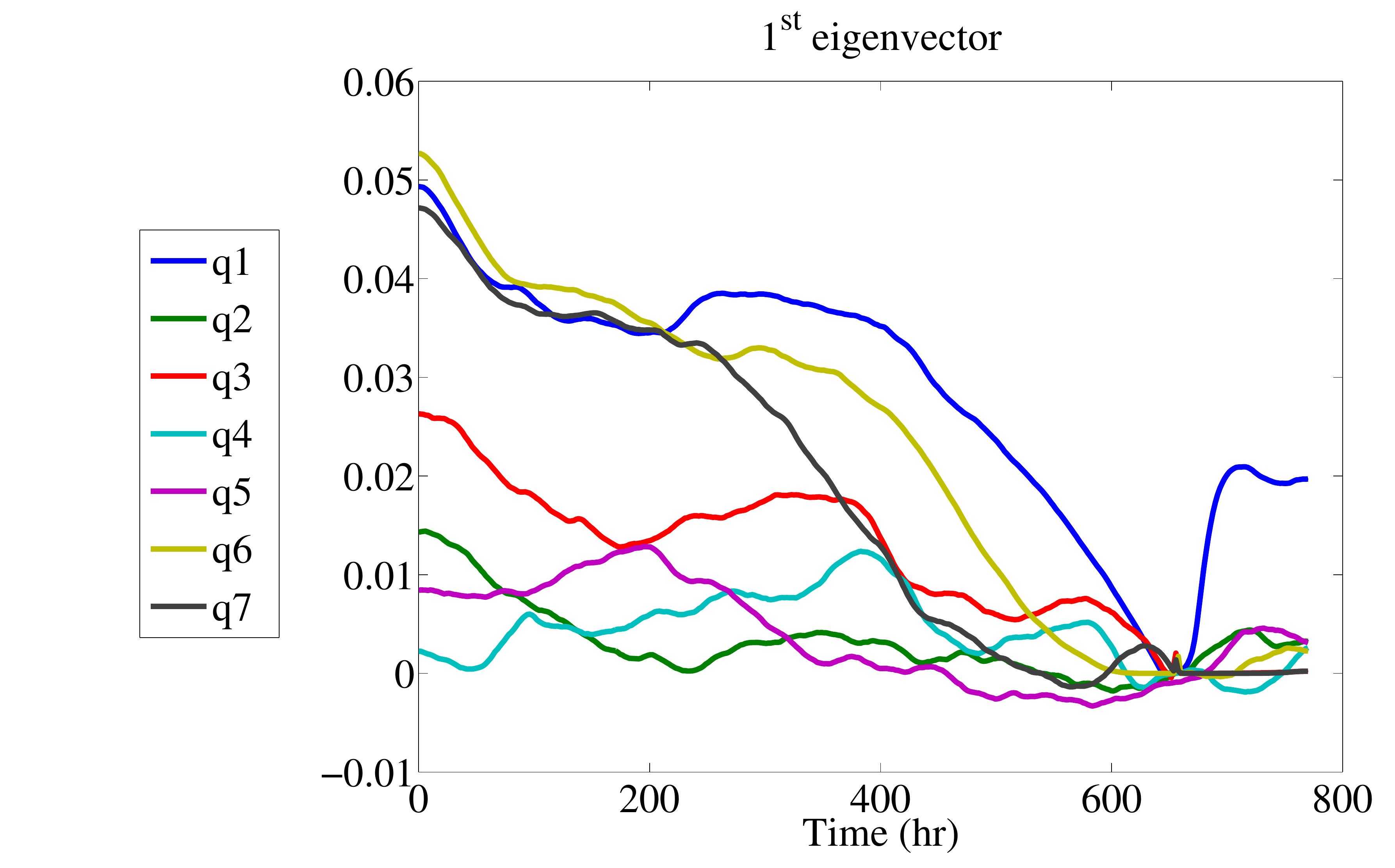}
    % \begin{subfigure}{0.40 \textwidth}
    %   \includegraphics[height=4cm, clip=true, trim=3cm 0cm 0cm 0cm]{./figs/constant}
    % \end{subfigure}
    & \raisebox{0.17\textheight}{b)}
    & \includegraphics[height=0.2\textheight, clip=true, trim=4cm 0cm 0cm 0cm]{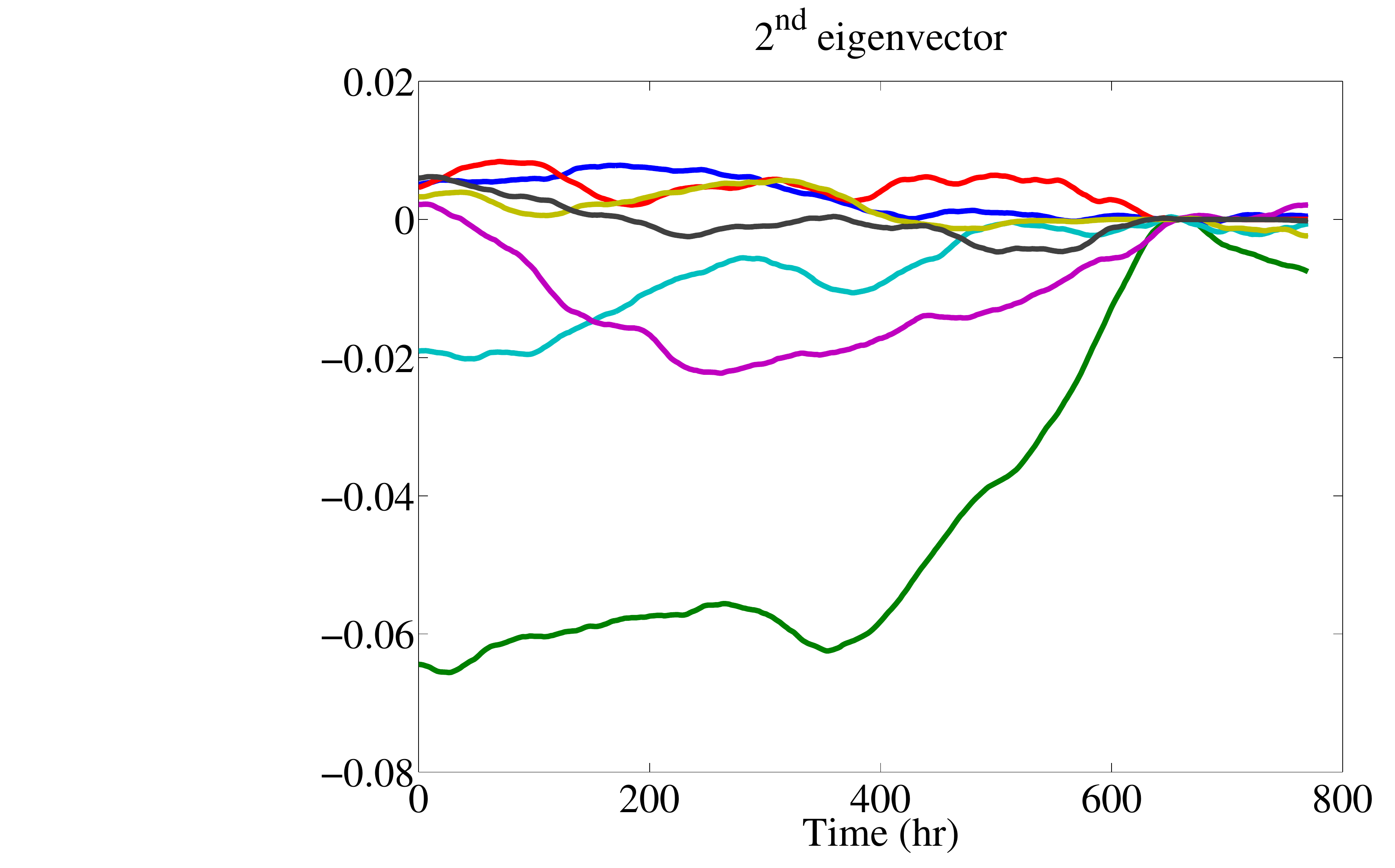}
    \\
    \raisebox{0.17\textheight}{c)}
    & \includegraphics[height=0.2\textheight, clip=true, trim=4cm 0cm 0cm 0cm]{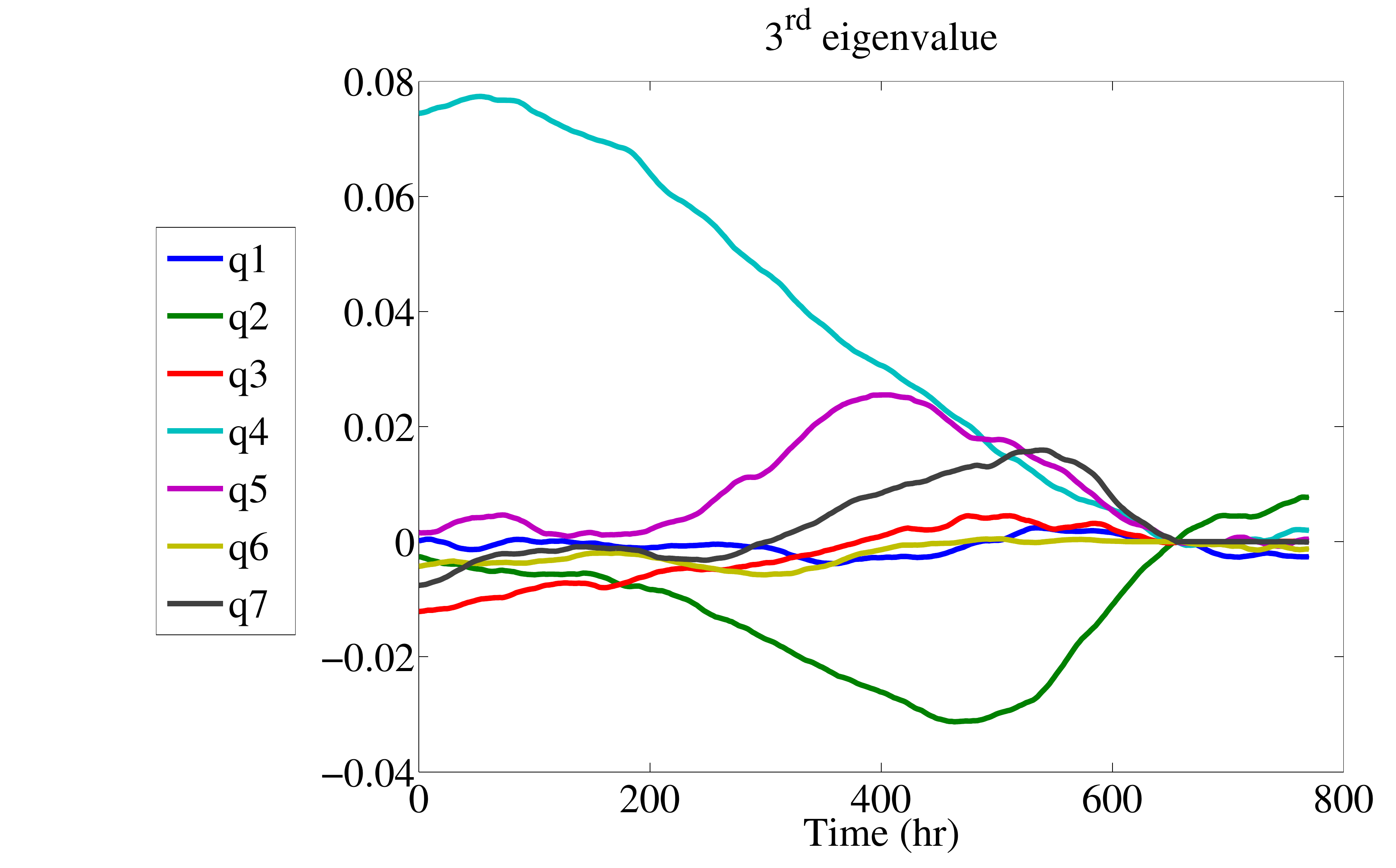}
    & \raisebox{0.17\textheight}{d)}
    & \includegraphics[height=0.2\textheight, clip=true, trim=8cm 0cm 0cm 0cm]{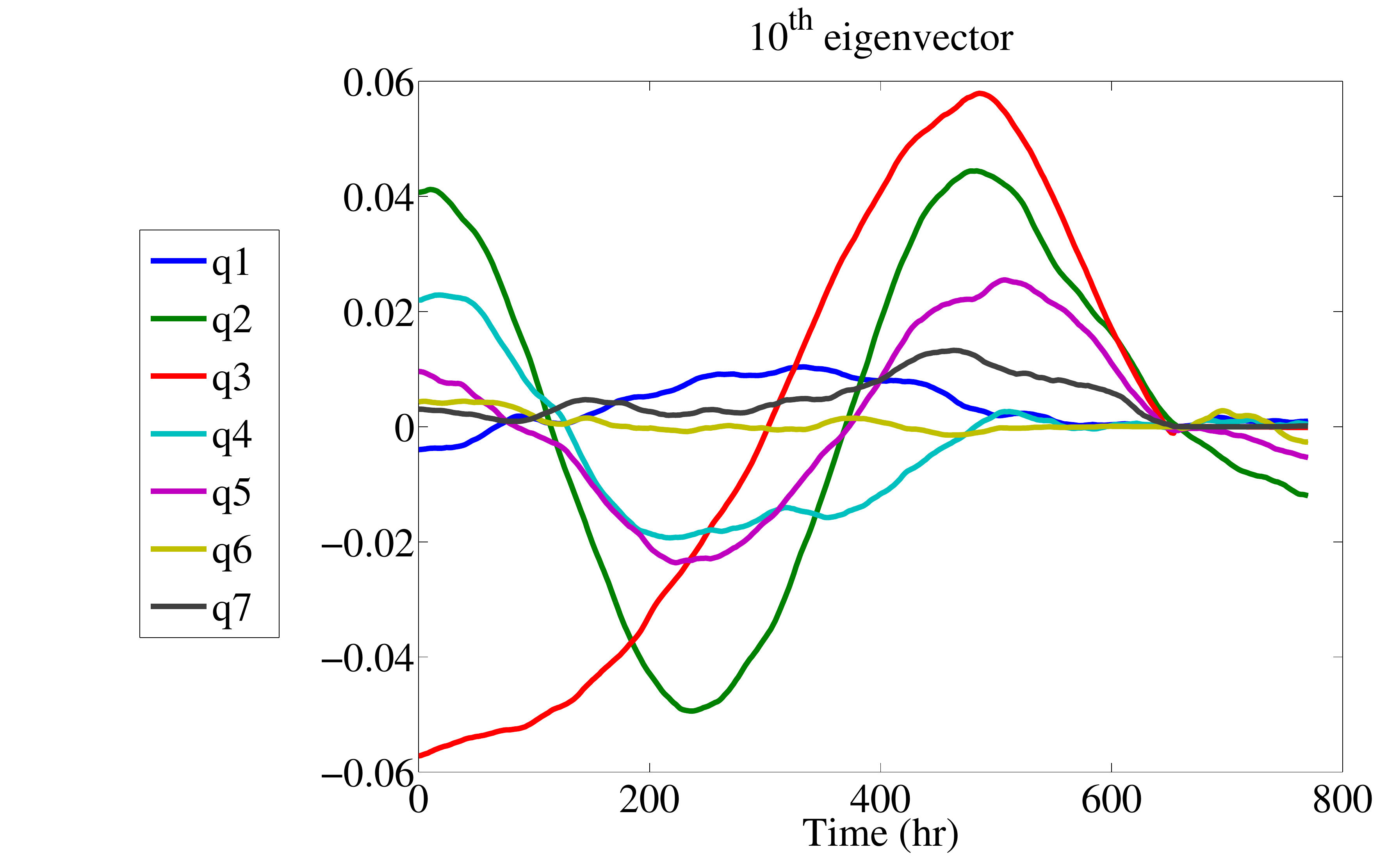}
    \\
    % \begin{subfigure}{0.40 \textwidth}
    %   \includegraphics[height=4cm, clip=true, trim=3cm 0cm 0cm 0cm]{./figs/constant}
    % \end{subfigure}
    \raisebox{0.17\textheight}{e)}
    & \includegraphics[height=0.2\textheight, clip=true, trim=4cm 0cm 0cm 0cm]{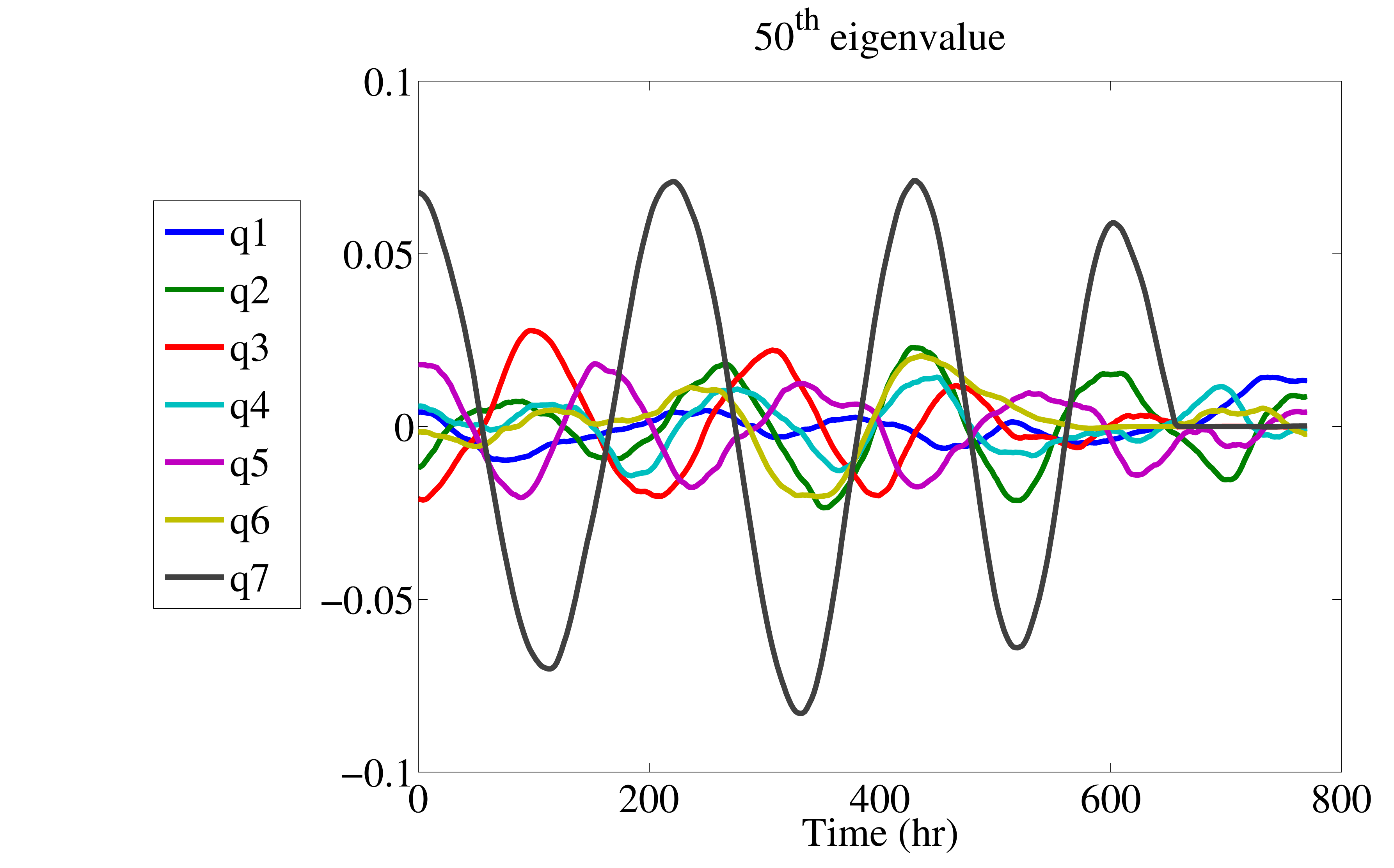}
    & \raisebox{0.17\textheight}{f)}
    & \includegraphics[height=0.2\textheight, clip=true, trim=4cm 0cm 0cm 0cm]{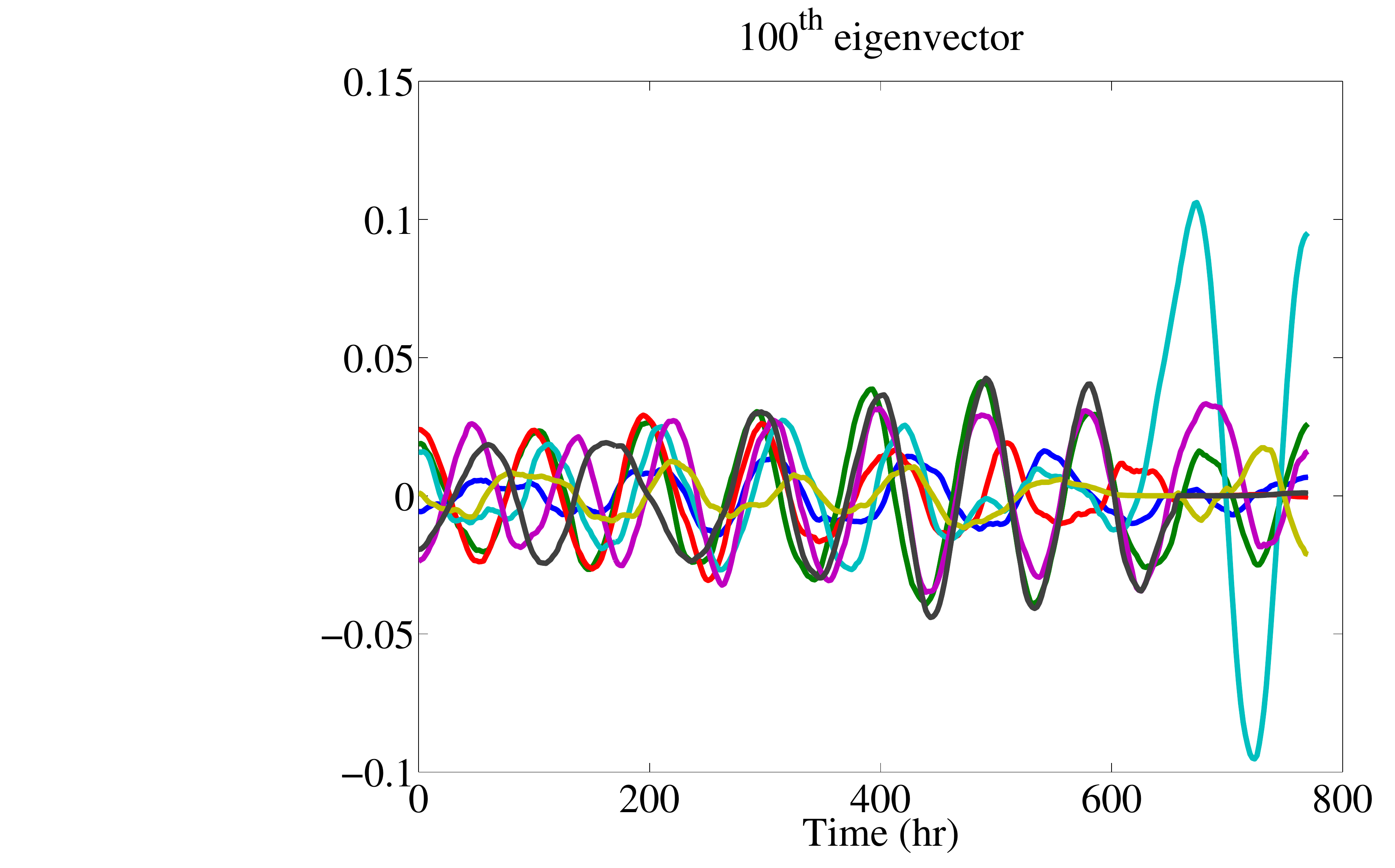}
  \end{tabular}
  \caption{Selected normalized eigenvectors of the posterior covariance
    for the solution with positivity constraint.  The first few
    eigenvalues correspond to directions of maximum uncertainty in the
    solution.}
  \label{fig:eigenvectors}
\end{figure}

In Figure~\ref{fig:eigenvectors} we depict a few selected eigenvectors
of the posterior covariance, where larger eigenvalues indicate
directions of higher uncertainty. Our low-rank approximation of the
covariance matrix preserves the eigen-directions and dismisses the more
oscillatory directions arising from smaller
eigenvalues. Figure~\ref{fig:forward-uq} shows the final result of our
uncertainty propagation study, with the standard deviation of the
depositions computed by retaining $n_e=100$ eigenvectors of
$\mb{C}_{\text{sp}}$, and presented on the same $100 \times 100$ spatial
grid that was used in Figure~\ref{fig:forward-uq}a. The contours of
standard deviation indicate the spatial variation in the uncertainty of
the deposition estimates around the industrial site.  It is clear from
Figure~\ref{fig:forward-uq}b that the deposition estimates become more
uncertain closer to the sources. However, uncertainties are very small
outside the boundaries of the smelter site, which means that we can be
confident in our estimate of the impact of emissions on the surrounding
area, even though we observed relatively large uncertainties in
estimates of the emission rates in Section~\ref{sec:real-data}.

\section{Conclusions}

In this paper we studied the inverse problem corresponding to estimating
the rate of fugitive emissions for airborne particulate matter from an
industrial site. We restrict ourselves to short-range transport of
particles (on the order of a few kilometres from the source) and make
simplifying assumptions that allow us to use an efficient Gaussian plume
type solution as our forward solver.  We then model our measurement
devices and construct a linear forward solver.

We solved the inverse problem using a Bayesian framework, developing a
solution approach that employs three possible prior assumptions, based
on the emission rates being: constant and positive; smoothly-varying but
not necessarily positive; and smoothly-varying and positive.  The
solution for these three cases becomes successively more flexible, but
also more expensive to compute.  We discuss methods for efficient
solution of the inverse problem in each case.  Finally, we apply our
solution framework to a concrete study of fugitive emissions of lead
particulates from a lead-zinc smelter in Trail, British Columbia,
Canada. We obtain estimates of emissions from seven suspect area sources
contained within the boundaries of the industrial site and then
extrapolate our solution to obtain a total annual emission rate.
Our estimates are consistent with the results of a independent study
that was performed by the company in 2013 \cite{thep-report}, which gives us a high degree of confidence in our choice of priors and the solution methodology.

There are several obvious areas for future research that aim to extend
the framework developed in this study. For example, one can consider
estimating emission rates for more than just one particulate.  In many
applications certain tracer particulates can be connected with specific
operations or sources, so that incorporating these tracers within the
framework has the potential to improve estimates of the emission rates
by distinguishing depositions that originate from specific
sources. Another interesting problem is that of identifying a singular
emission event, or in other words estimating emissions from a source
that emits a large amount of particulate material during a relatively
short time period. In this case, one would have to consider a different
prior that can incorporate singular sources in contrast with the
smoothness priors used in this study.  Finally, the optimal experimental
design for source inversion problems is of great interest in practical
applications. For example, determining optimal locations for dust-fall
jars and real-time measurement devices could significantly improve both
the data quality and the emission estimates.  It is crucial to have a
good strategy for deploying sensors, and this strategy will be subject
to practical constraints and may depend on factors such as wind patterns
that are not known at the time of the measurements. One promising
approach is to apply statistical learning techniques to predict unknown
parameters based on existing data, which is related to our method in
Section~\ref{sec:params} for regularizing and extending wind data using
Gaussian processes.

% Finally, we studied the implied impact of our estimates of the emission
% rates on the area surrounding the industrial site by computing the
% ground level depositions. We also discussed a framework for propagation
% of posterior uncertainty through the forward model to estimate the
% uncertainty of the computed deposition values. The cornerstone of our
% approach here was to exploit the fact that the posterior had rapidly
% decaying eigenvalues in order to construct a low-rank approximation of
% the posterior covariance that can be propagated through the forward
% model at low cost.

% \todo{[I think you could drop this last paragraph.  It doesn't add much.
%   Instead, I think it's a good idea to conclude with a few ideas for
%   possible extensions of this method and areas for future research.]}

\section*{Acknowledgements}

We would like to thank Peter Golden, Cheryl Darrah and Mark Tinholt from
Teck's Trail Operations for many valuable discussions.  This project was
supported by an NSERC Discovery Grant (to JMS) and an Accelerate
Internship Grant from Mitacs and Teck Resources (to BH).

%\section*{References}
\bibliographystyle{abbrv}   %% dash under repeated name, von ignored
%%\addcontentsline{toc}{chapter}{Bibliography}
%%\typeout{Bibliography}
\bibliography{references}%,atmos}

\end{document}